\documentclass[journal]{IEEEtran}
\IEEEoverridecommandlockouts

\usepackage{cite}
\usepackage{amsmath,amssymb,amsfonts}
\usepackage{algorithmic}
\usepackage{graphicx}
\usepackage{textcomp}
\usepackage[usenames,dvipsnames,table]{xcolor}

\usepackage{amsthm} 
\usepackage{multirow}
\usepackage[overload]{empheq}
\usepackage{xspace} 
\usepackage{marvosym} 
\usepackage{hyperref}
    \hypersetup{colorlinks=true,linkcolor=black,citecolor=black,urlcolor=black}
\usepackage{todonotes} 
\usepackage{ifthen} 
\usepackage{cleveref} 
\crefname{subsection}{subsection}{subsections}
\crefname{subsubsection}{subsubsection}{subsubsections}
\crefname{algorithm}{algorithm}{algorithms}

\def\BibTeX{{\rm B\kern-.05em{\sc i\kern-.025em b}\kern-.08em
    T\kern-.1667em\lower.7ex\hbox{E}\kern-.125emX}}

\newcommand{\code}[1]{\texttt{#1}}
\newcommand{\TheTeam}{TAMIS\xspace}
\newcommand{\TheProject}{TeamPlay\xspace}

\newcommand{\pref}[1]{p.\pageref{#1}}

\newcommand{\Crpref}[1]{\Cref{#1} \pref{#1}}

\newcommand{\todoColor}{\todoColorDefault}
\newcommand{\todoSetColor}[1]{
    \ifthenelse{\equal{#1}{YM}}{
        \renewcommand{\todoColor}{BurntOrange}}{}%
    \ifthenelse{\equal{#1}{TR}}{
        \renewcommand{\todoColor}{Bittersweet}}{}%
    \ifthenelse{\equal{#1}{PV}}{
        \renewcommand{\todoColor}{ForestGreen}}{}%
}

\DeclareMathOperator{\public}{\text{public}}
\DeclareMathOperator{\secret}{\text{secret}}
\newcommand{\init}{\text{init}}
\newcommand{\bit}[2]{#1\!\left[#2\right]}
\newcommand{\iInit}{1}
\newcommand{\iFinal}{n}
\newcommand{\rand}{\text{random}}
\newcommand{\egcd}{\text{EEA}}

\newcommand{\knowing}{\ \middle|\ }
\newcommand{\proba}[2]{\ifthenelse{\equal{\unexpanded{#1}}{}}{Pr}{#1}\!\left(#2\right)}

\newcommand{\Z}{\mathbb{Z}}
\newcommand{\F}{\mathbb{F}}
\newcommand{\set}[2]{\left\{{#1}\ifthenelse{\equal{\unexpanded{#2}}{}}{}{\knowing {#2}}\right\}}
\newcommand{\card}[1]{\text{card}\!\left({#1}\right)}

\newcommand{\floor}[1]{\left\lfloor{#1}\right\rfloor}
\newcommand{\pgcd}[2]{\gcd\!\left({#1}, {#2}\right)}
\newcommand{\ZnZ}[1]{\Z/{#1}\Z}
\newcommand{\ZnZinv}[1]{\left(\ZnZ{#1}\right)^{\star}}
\newcommand{\Field}[1]{\F_{#1}}

\newcommand{\symbThen}{\theta}
\newcommand{\symbElse}{\varepsilon}
\newcommand{\symbFirst}{f}
\newcommand{\symbSecond}{g}
\newcommand{\symbLadder}{\ell}
\newcommand{\funThen}[1]{\symbThen\left({#1}\right)}
\newcommand{\funElse}[1]{\symbElse\left({#1}\right)}
\newcommand{\funFirst}[2]{\symbFirst\left({#1},{#2}\right)}
\newcommand{\funSecond}[2]{\symbSecond\left({#1},{#2}\right)}
\newcommand{\funLadder}[1]{\symbLadder\left({#1}\right)}

\newcommand{\cThenA}{a}

\newcommand{\cFirstA}{\symbFirst_{20}}
\newcommand{\cFirstB}{\symbFirst_{11}}
\newcommand{\cFirstC}{\symbFirst_{02}}
\newcommand{\cFirstD}{\symbFirst_{10}}
\newcommand{\cFirstE}{\symbFirst_{01}}
\newcommand{\cFirstF}{\symbFirst_{00}}
\newcommand{\cSecondA}{\symbSecond_{20}}
\newcommand{\cSecondB}{\symbSecond_{11}}
\newcommand{\cSecondC}{\symbSecond_{02}}
\newcommand{\cSecondD}{\symbSecond_{10}}
\newcommand{\cSecondE}{\symbSecond_{01}}
\newcommand{\cSecondF}{\symbSecond_{00}}
\newcommand{\cLadderA}{\symbLadder_{2}}
\newcommand{\cLadderB}{\symbLadder_{1}}
\newcommand{\cLadderC}{\symbLadder_{0}}

\newcommand{\cWeakB}{m}
\newcommand{\cStrong}{c}
\newcommand{\cStrongA}{\cStrong_{0}}
\newcommand{\cStrongB}{\cStrong_{1}}
\newcommand{\cStrongC}{\cStrong_{2}}
\newcommand{\cStrongD}{\cStrong_{3}}
\newcommand{\cVal}{v}
\newcommand{\cValA}{\cVal_{0}}
\newcommand{\cValC}{\cVal_{2}}
\newcommand{\cValD}{\cVal_{3}}
\newcommand{\cInverse}{u}
\newcommand{\cInverseB}{\cInverse_{1}}
\newcommand{\cInverseC}{\cInverse_{2}}
\newcommand{\cInverseD}{\cInverse_{3}}
\newcommand{\cGcd}{d}
\newcommand{\cGcdB}{\cGcd_{1}}
\newcommand{\cGcdC}{\cGcd_{2}}
\newcommand{\cGcdD}{\cGcd_{3}}

\newcommand{\FI}[1]{\text{\Lightning}{#1}}
\newcommand{\xInit}{x_\text{init}}
\newcommand{\yInit}{y_\text{init}}
\newcommand{\xFinal}{x_\text{final}}
\newcommand{\yFinal}{y_\text{final}}
\newcommand{\xFault}{x_\text{fault}}
\newcommand{\yFault}{y_\text{fault}}
\newcommand{\run}{\text{EXE}}

\newcommand{\nBits}{\text{nBits}}
\newcommand{\intervalLadder}[2]{\text{L}_{{#1}, {#2}}}
\newcommand{\condN}[1]{\raisebox{.5pt}{\textcircled{\raisebox{-.9pt} {{#1}}}}}
\newcommand{\residue}[2]{\text{R}^3_{#1}\!\left({#2}\right)}

\newcommand{\pointO}{\mathcal{O}}

\newcommand{\pointA}{A}
\newcommand{\pointP}{P}
\newcommand{\pointQ}{Q}
\newcommand{\pointR}{R}
\newcommand{\cFirstPointP}{\symbFirst_{\pointP}}
\newcommand{\cFirstPointQ}{\symbFirst_{\pointQ}}
\newcommand{\cFirstPointA}{\symbFirst_{\pointA}}
\newcommand{\cSecondPointP}{\symbSecond_{\pointP}}
\newcommand{\cSecondPointQ}{\symbSecond_{\pointQ}}
\newcommand{\cSecondPointA}{\symbSecond_{\pointA}}
\newcommand{\cLadderPointP}{\symbLadder_{\pointP}}
\newcommand{\cLadderPointA}{\symbLadder_{\pointA}}
\newcommand{\cEndormophism}{\lambda}

\theoremstyle{plain}
\newtheorem{theo}{Theorem}
\newtheorem{lem}{Lemma}

\newtheorem*{lem*}{Lemma}
\newtheorem*{cor*}{Corollary}
\newtheorem*{prop*}{Proposition}

\theoremstyle{definition}
\newtheorem{defi}{Definition}

\theoremstyle{remark}

\begin{document}

\title{A Hole in the Ladder:\\
{\huge Interleaved Variables in Iterative Conditional Branching (Extended Version)}
\thanks{This work was supported by the EU Horizon 2020 project \emph{TeamPlay} \mbox{(\url{https://www.teamplay-h2020.eu}}), grant number 779882.}
}

\author{\IEEEauthorblockN{
    Yoann Marquer\IEEEauthorrefmark{1},
    Tania Richmond\IEEEauthorrefmark{2},
    Pascal V\'eron\IEEEauthorrefmark{3}
}
\IEEEauthorblockA{\\
    \IEEEauthorrefmark{1}\textit{\TheTeam team then DiverSE team, \TheProject project, Inria, Univ. Rennes, CNRS, IRISA, France}, 
        yoann.marquer@inria.fr,\\
    \IEEEauthorrefmark{2}\textit{\TheTeam team, \TheProject project Inria, Univ. Rennes, CNRS, IRISA, France}, then
        \textit{DGA - Ma\^{i}trise de l’Information, Bruz, France},
        tania.richmond@inria.fr,\\
    \IEEEauthorrefmark{3}\textit{Laboratoire IMath, Université de Toulon, France},
        pascal.veron@univ-tln.fr}
}

\maketitle

\begin{abstract}
The iterative conditional branchings appear in va\-rious sensitive algorithms, like the modular exponentiation in the RSA cryptosystem or the scalar multiplication in elliptic-curve cryptography.
In this paper, we abstract away the desirable security properties achieved by the Montgomery ladder, and formalize systems of equations necessary to obtain what we call the semi-interleaved and fully-interleaved ladder properties.
This fruitful approach allows us to design novel fault-injection attacks, able to obtain some/all bits of the secret against different ladders, including the common Montgomery ladder.
We also demonstrate the generality of our approach by applying the ladder equations to the modular exponentiation and the scalar multiplication, both in the semi- and fully-interleaved cases, thus proposing novel and more secure algorithms.
\end{abstract}


\section{Introduction}
\label{sec:intro}

We present common algorithms used to compute the modu\-lar exponentiation, based on an \emph{iterative conditional bran\-ching}, where a conditional branching depending on the secret updates the value of a variable $x$ on every iteration of one/several loop(s).
Amongst these algorithms the Montgomery ladder, that uses a fresh variable $y$, satisfies desirable properties against side-channel or fault-injection attacks.

\subsection{Contribution}
\label{sec:intro:contribution}
In this paper, we formalize these properties with equations corresponding to two families of cases: the \emph{semi-interleaved ladders} where the value of $x$ or $y$ may (depending on the secret) depend only on its previous value, and the \emph{fully-interleaved ladders} where the value of $x$ and $y$ both depend on both previous values.

We propose 1) an attacker model using fault-injection able to obtain some bits of the secret in the semi-interleaved cases (including the Montgomery ladder) but none in the fully-interleaved cases, and 2) a stronger attacker model able to obtain all bits of the secret in the semi-interleaved cases, and even in the fully-interleaved cases if the fault can be injected in the key register.

We also use our formalization to design novel algorithms for the modular exponentiation, including a semi-interleaved ladder with a mask updated at every iteration, and a fully-interleaved ladder.
We demonstrate the generality of the approach by also applying the ladder equations to the scalar multiplication used in elliptic-curve cryptography (ECC), obtaining a semi-interleaved ladder suitable for practical applications, and a fully-interleaved ladder.

The ladder equations, the first two attacks and the algorithms for the modular exponentiation have been published at ARITH 2020 \cite{MR20}.
This paper is an extended version, including novelties like:
\begin{itemize}
        \item the third attack (for the second attacker model) against all (including fully-interleaved) ladders (\Cref{sec:faultInjection:attacker2}),
        \item in the case of the modular exponentiation a proof that the probability to obtain randomly a suitable ladder constant for our fully-interleaved ladder candidate is almost $1$ when used in concrete applications like RSA or DSA cryptosystems (\Cref{sec:exponentiation:RSA,sec:exponentiation:DSA}),
        \item and original semi- and fully-interleaved ladder candidates for the scalar multiplication over elliptic curves (\Cref{sec:scalarMult}).
\end{itemize}


\subsection{Organization of the Paper}
\label{sec:intro:organization}

As introduction, we present in \Cref{sec:soa} some related works on the modular exponentiation and the Montgomery ladder.
In \Cref{sec:ladderEquations} we formalize the iterative conditional branching to deduce the equations satisfied by the semi-interleaved and fully-interleaved ladders, and thus the requirements for ladderizable algorithms.
In \Cref{sec:faultInjection} we introduce two attacker models using fault injection techniques, and compare the vulnerability of the non-, semi- and fully-interleaved ladders.
Finally, we detail how to produce examples of the semi- and fully-interleaved ladders in \Cref{sec:exponentiation} for the modular exponentiation and in \Cref{sec:scalarMult} for the scalar multiplication.

\section{Related Works}
\label{sec:soa}

In this section, we introduce known algorithms for the modular exponentiation, their relevance regarding security, and the desirable properties of the usual Montgomery ladder.

\subsection{Modular Exponentiation}
\label{sec:soa:exp}

Let $k$ be a secret key, and $k = \sum_{0 \le i \le d} \bit{k}{i}2^i$ be its binary expansion of size $d+1$, i.e. $\bit{k}{i}$ is the bit $i$ of $k$.
The \emph{square-and-multiply} algorithm described in \Cref{table:squareMultiply} computes the (left-to-right) modular exponentiation $a^k \bmod n$, by using $a^{\sum_{0 \le i \le d} \bit{k}{i}2^i} = \prod_{0 \le i \le d} (a^{2^i})^{\bit{k}{i}}$.
This exponentiation is commonly used in cryptosystems like RSA~\cite{RSA78}.

\begin{table}[t]
    \begin{minipage}{0.49\linewidth}
        \begin{algorithmic}[1]
            \REQUIRE $\public a, n; \secret k$
            \STATE $x \leftarrow 1$
            \FOR{$i = d$ \TO $0$}
                \STATE $x \leftarrow x^2 \bmod n$
                \IF{$\bit{k}{i} = 1$}
                    \STATE $x \leftarrow a x \bmod n$
                \ENDIF
            \ENDFOR
            \RETURN $x$
            \ENSURE $x = a^k \bmod n$
        \end{algorithmic}
        \caption{Square and Multiply}
        \label{table:squareMultiply}
    \end{minipage}
    \hfill
    \begin{minipage}{0.49\linewidth}
        \begin{algorithmic}[1]
            \REQUIRE $\public a, n; \secret k$
            \STATE $x \leftarrow 1$
            \FOR{$i = d$ \TO $0$}
                \STATE $x \leftarrow x^2 \bmod n$
                \IF{$\bit{k}{i} = 1$}
                    \STATE $x \leftarrow a x \bmod n$
                \ELSE
                    \STATE $y \leftarrow a x \bmod n$
                \ENDIF
            \ENDFOR
            \RETURN $x$
            \ENSURE $x = a^k \bmod n$
        \end{algorithmic}
        \caption{Square and Multiply Always}
        \label{table:squareMultiplyAlways}
    \end{minipage}
\end{table}

For every iteration, the multiplication $ax$ is computed only if $\bit{k}{i} = 1$, which can be detected by observing execution time%
\footnote{Even observing the duration of the whole execution can leak the Hamming weight of the secret key, and thus can narrow down the exploration space.}
\cite{Koc96} or power profiles%
\footnote{A multiplication can be distinguished from a squaring, hence variants with squaring only have been proposed in \cite{CFG+11} and improved in \cite{HCC+14}.}
by means of e.g. SPA (Simple Power Analysis%
\footnote{The power profile depends also on the values in the considered registers, so computing a multiplication in every case is better against SPA but not against CPA (Correlation Power Analysis) \cite{BCO04}, even if standard blinding techniques can prevent differential attacks \cite{Cor99,MDS99}.}%
) \cite{KJJ99}, and thus leads to information leakage from both time and power side-channel attacks.

To prevent SPA, regularity of the modular exponentiation algorithms is required, which means that both branches of the sensitive conditional branching perform the same operations, independently from the value of the exponent.
Thus, an \code{else} branch is added with a dummy instruction \cite{Cor99} in the \emph{square-and-multiply-always} algorithm described in \Cref{table:squareMultiplyAlways}.

But countermeasures developed against a given attack may benefit another one \cite{SKL+02}.
Because the multiplication in the \code{else} branch of this algorithm is a dummy operation, a fault injected \cite{ZAV04} in the register containing $ax$ will eventually propagate through successive iterations and alter the final result only if $\bit{k}{i} = 1$, thus leaking some information.
Therefore, an attacker (see the attacker models in \Cref{sec:faultInjection}) able to inject a fault in a given register at a given iteration can obtain the digits of the secret key by comparing the final output with or without fault (technique known as safe-error attack).

\subsection{Montgomery Ladder}
\label{sec:soa:montgomery}

This is not the case in the algorithm proposed by Montgomery \cite{Mon87} and described in \Cref{table:montgomeryLadder}, where a fault injected in a register will eventually propagate to the other one, and thus will alter the final result in any case, preventing the attacker to obtain information.
But, as described in \Cref{sec:faultInjection}, some information on the last digits may still leak, weakening the protection obtained from the ladder.

\begin{table}[t]
    \begin{minipage}{0.49\linewidth}
        \begin{algorithmic}[1]
            \REQUIRE $\public a, n; \secret k$
            \STATE $x \leftarrow 1$
            \STATE $y \leftarrow a \bmod n$
            \FOR{$i = d$ \TO $0$}
                \IF{$\bit{k}{i} = 1$}
                    \STATE $x \leftarrow x y \bmod n$
                    \STATE $y \leftarrow y^2 \bmod n$
                \ELSE
                    \STATE $y \leftarrow x y \bmod n$
                    \STATE $x \leftarrow x^2 \bmod n$
                \ENDIF
            \ENDFOR
            \RETURN $x$
            \ENSURE $x = a^k \bmod n$
        \end{algorithmic}
        \caption{Montgomery Ladder}
        \label{table:montgomeryLadder}
    \end{minipage}
    \hfill
    \begin{minipage}{0.49\linewidth}
        \begin{algorithmic}[1]
            \STATE $x \leftarrow \init$
            \FOR{$i = \iInit$ \TO $\iFinal$}
                \STATE $\ddots$
                \IF{$\secret$}
                    \STATE $x \leftarrow \funThen{x}$
                \ELSE
                    \STATE $x \leftarrow \funElse{x}$
                \ENDIF
                \STATE \reflectbox{$\ddots$}
            \ENDFOR
        \end{algorithmic}
        \caption{Iterative Conditional Branching}
        \label{table:iterCondBranch}
    \end{minipage}
\end{table}

The Montgomery ladder is algorithmically equivalent \cite{Mar19} to the square-and-multiply(-always) algorithm(s), in the sense that $x$ has the same value for every iteration.
Actually, some variants \cite{BNP07} of the square-and-multiply-always algorithm%
\footnote{See \cite{Joy07} for highly regular right-to-left variants, \cite{Joy09} for a generalization to any basis and left-to-right/right-to-left variants, and \cite{Wal17} for their duality.}
may be as resistant as the Montgomery ladder \cite{JY+03}, both by checking invariants \cite{KKR+16} violated if a fault is injected.
In the case of the Montgomery ladder, the invariant $y = ax$ is satisfied for every iteration.
These invariants are important for the self-secure exponentiation countermeasures \cite{Gir06}.

Note that the \code{else} branch in \Cref{table:montgomeryLadder} is identical to the \code{then} branch, except that $x$ and $y$ are swapped, which provides also (partial%
\footnote{The variables $x$ and $y$ may have different access time, which hinders protection against cache-timing leakage.}%
) protection against timing and power leakage.
Moreover, the variable dependency makes these variables interleaved, so this exponentiation is algorithmically (but partially, as we will demonstrate) protected against safe-error attacks.
Finally, as opposed to square-and-multiply-always in \Cref{table:squareMultiplyAlways},
the code in the \code{else} branch in \Cref{table:montgomeryLadder} is not dead, so will not be removed by compiler optimizations.

\section{Ladder equations}
\label{sec:ladderEquations}

In this section, we formalize the iterative conditional branching occuring in algorithms like the Montgomery ladder, and deduce the requirements to optimize them with semi- or fully-interleaved ladders.

\subsection{Iterative Conditional Branching}
\label{sec:ladderEquations:iterCondBranch}

In this paper, we focus on algorithms as in \Cref{table:iterCondBranch} called \emph{iterative conditional branching}.
It appears in algorithms like square-and-multiply for the modular exponentiation, its counterpart double-and-add for elliptic curve point multiplication \cite{Mil85,Kob87}, or the secure bit permutation in the McEliece cryptosystem \cite{STM+08} attacked in \cite{PRD+16}.
It also appears naturally (\Cref{table:loopSecret}) when trying to turn a loop on the secret into a conditional branching depending on the secret, that can itself be balanced to remove or reduce the dependency on the secret.

\begin{table}[t]
    \begin{minipage}{0.49\linewidth}
        \begin{algorithmic}[1]
            \STATE $\text{assert}(\text{secret} \le \text{bound})$
            \FOR{$i = 0$ \TO $\text{secret}$}
                \STATE \text{\vdots}
            \ENDFOR
        \end{algorithmic}
    \end{minipage}
    \hfill
    \begin{minipage}{0.49\linewidth}
        \begin{algorithmic}[1]
            \FOR{$i = 0$ \TO $\text{bound}$}
                \IF{$i \le \text{secret}$}
                    \STATE \text{\vdots}
                \ENDIF
            \ENDFOR
        \end{algorithmic}
    \end{minipage}
    \vspace{0.2cm}
    \caption{Loop bounded by a sensitive variable}
    \label{table:loopSecret}
\end{table}

But our approach does not depend on the number/depth of the considered loops, hence the dots in \Cref{table:iterCondBranch}.
We assume only that the conditional branching uses only one variable $x$, the multivariate case being future work (see \Cref{sec:conclu:futureWork}).

\begin{defi}[Iterative Conditional Branching]
\label{defi:iterCondBranch}
An algorithm as in \Cref{table:iterCondBranch} is said with an (univariate) \emph{iterative conditional branching} with two (unary) functions $\symbThen$ and $\symbElse$.
\end{defi}

\subsection{Semi-Interleaved Ladders}
\label{sec:ladderEquations:weakLadder}

\begin{table}[t]
    \begin{minipage}{0.49\linewidth}
        \begin{algorithmic}[1]
            \STATE $x \leftarrow \init$
            \STATE $y \leftarrow \funLadder{\init}$
            \FOR{$i = \iInit$ \TO $\iFinal$}
                \STATE \dots
                \IF{$\secret$}
                    \STATE $x \leftarrow \funFirst{x}{y}$
                    \STATE $y \leftarrow \funElse{y}$
                \ELSE
                    \STATE $y \leftarrow \funFirst{y}{x}$
                    \STATE $x \leftarrow \funElse{x}$
                \ENDIF
                \STATE \dots
            \ENDFOR
        \end{algorithmic}
        \caption{Semi-Interleaved ladders}
        \label{table:weakLadder}
    \end{minipage}
    \hfill
    \begin{minipage}{0.49\linewidth}
        \begin{algorithmic}[1]
            \STATE $x \leftarrow \init$
            \STATE $y \leftarrow \funLadder{\init}$
            \FOR{$i = \iInit$ \TO $\iFinal$}
                \STATE \dots
                \IF{$\secret$}
                    \STATE $x \leftarrow \funFirst{x}{y}$
                    \STATE $y \leftarrow \funSecond{x}{y}$
                \ELSE
                    \STATE $y \leftarrow \funFirst{y}{x}$
                    \STATE $x \leftarrow \funSecond{y}{x}$
                \ENDIF
                \STATE \dots
            \ENDFOR
        \end{algorithmic}
        \caption{Fully-Interleaved ladders}
        \label{table:strongLadder}
    \end{minipage}
\end{table}

To prevent information leakage from side-channel analysis or fault injections, we use another variable $y$ in the algorithm described in \Cref{table:weakLadder}.
As in the Montgomery ladder (see \Cref{table:montgomeryLadder}),
we need to find two functions $\symbLadder$ and $\symbFirst$ such that for every iteration:
\begin{itemize}
    \item $y = \funLadder{x}$, and
    \item $x$ has the same value for every iteration as in \Cref{table:iterCondBranch}.
\end{itemize}
$y = \funLadder{x}$ is satisfied at the initialization.
By induction, let's assume $y = \funLadder{x}$ at the beginning of an iteration.
In the \code{then} branch%
\footnote{Because the condition depends on the secret, we will not assume that in the \code{then} branch the condition is satisifed but not in the \code{else} branch, to prevent data dependencies that could be attacked.}
we have $x \leftarrow \funFirst{x}{y}$ then $y \leftarrow \funElse{y}$, thus in order to have $y = \funLadder{x}$ satisfied at the end of an iteration the following equation must hold:
\[\forall x, \funElse{\funLadder{x}} = \funLadder{\funFirst{x}{\funLadder{x}}}\]
and 
to have $x$ updated to $\funThen{x}$ during the iteration the following equation must hold:
\[\forall x, \funFirst{x}{\funLadder{x}} = \funThen{x}\]
In the \code{else} branch we have $y \leftarrow \funFirst{y}{x}$ then $x \leftarrow \funElse{x}$, thus in order to have $y = \funLadder{x}$ satisfied at the end of an iteration the following equation must hold:
\[\forall x, \funFirst{\funLadder{x}}{x} = \funLadder{\funElse{x}}\]
and $x$ is already updated to $\funElse{x}$ during the iteration.

\begin{defi}[Semi-Ladderizable]
\label{defi:weakLadder}
Let $A$ be an algorithm with a univariate iterative conditional branching with two unary functions denoted $\symbThen$ and $\symbElse$. $A$ is \emph{semi-ladderizable} if there exists a unary function $\symbLadder$ and a binary function $\symbFirst$ such that, for every considered value $x$:
\begin{align}[left=\empheqlbrace]
    \funElse{\funLadder{x}} &= \funLadder{\funThen{x}}\label{eq:weak:thenLadder}\\
    \funFirst{x}{\funLadder{x}} &= \funThen{x}\label{eq:weak:thenValue}\\
    \funFirst{\funLadder{x}}{x} &= \funLadder{\funElse{x}}\label{eq:weak:elseLadder}
\end{align}
\end{defi}

For the square-and-multiply algorithm 
we have $\funThen{x} = ax^2$ and $\funElse{x} = x^2$, and we know that it can be semi-ladderized by using the Montgomery ladder 
with $\funLadder{x} = ax$ and $\funFirst{x}{y} = xy$, but we demonstrate in \Cref{sec:exponentiation} that there are other solutions.
Note that to respect the form of the semi-interleaved ladder, we should have written
$y \leftarrow y x$ and not
$y \leftarrow x y$ in the \code{else} branch%
\footnote{The former is actually better regarding vulnerability to the M safe-error \cite{JY+03} or collision\cite{KKY+10} attacks, showing that our methodology is good practice.}
of the Montgomery ladder.

\subsection{Fully-Interleaved Ladders}
\label{sec:ladderEquations:strongLadder}

Unfortunately, the semi-interleaved ladder 
is vulnerable to fault injection techniques, because in every branch at least one variable depends only on its previous value and not the previous value of both variables (see \Cref{sec:faultInjection}).
Moreover, an attacker able to determine whether the output of one operation is used as the input to another one can \cite{HKT+15} apply collision attacks%
\footnote{A countermeasure proposed in \cite{LTT15} is to randomly blend variants of the ladder, or compute the exponentiation by taking a random (bounded) walk.}
to deduce whether two following bits are the same.
To solve these issues, we propose in \Cref{table:strongLadder} a \emph{fully-interleaved  ladder} using three functions $\symbLadder$, $\symbFirst$ and $\symbSecond$.

As for the semi-interleaved ladders, $y = \funLadder{x}$ is satisfied at the initialization.
We assume again by induction that $y = \funLadder{x}$ at the beginning of an iteration.
In the \code{then} branch we have $x \leftarrow \funFirst{x}{y}$ then $y \leftarrow \funSecond{x}{y}$, thus in order to have $y = \funLadder{x}$ satisfied at the end of an iteration the following equation must hold:
\[\forall x, \funSecond{\funFirst{x}{\funLadder{x}}}{\funLadder{x}} = \funLadder{\funFirst{x}{\funLadder{x}}}\]
and 
to have $x$ updated to $\funThen{x}$ during the iteration the following equation must hold:
\[\forall x, \funFirst{x}{\funLadder{x}} = \funThen{x}\]
In the \code{else} branch we have $y \leftarrow \funFirst{y}{x}$ then $x \leftarrow \funSecond{y}{x}$, thus, in order to have $y = \funLadder{x}$ satisfied at the end of an iteration, the following equation must hold:
\[\forall x, \funFirst{\funLadder{x}}{x} = \funLadder{\funSecond{\funFirst{\funLadder{x}}{x}}{x}}\]
and 
to have $x$ updated to $\funElse{x}$ during the iteration the following equation must hold:
\[\forall x, \funSecond{\funFirst{\funLadder{x}}{x}}{x} = \funElse{x}\]

\begin{defi}[Fully-Ladderizable]
\label{defi:strongLadder}
Let $A$ be an algorithm with a univariate iterative conditional branching with two unary functions denoted $\symbThen$ and $\symbElse$. $A$ is \emph{fully-ladderizable} if there exists a unary function $\symbLadder$ and two binary functions $\symbFirst$ and $\symbSecond$ such that, for every considered input value $x$:
\begin{align}[left=\empheqlbrace]
    \funSecond{\funThen{x}}{\funLadder{x}} &= \funLadder{\funThen{x}}\label{eq:strong:thenLadder}\\
    \funFirst{x}{\funLadder{x}} &= \funThen{x}\label{eq:strong:thenValue}\\
    \funFirst{\funLadder{x}}{x} &= \funLadder{\funElse{x}}\label{eq:strong:elseLadder}\\
    \funSecond{\funFirst{\funLadder{x}}{x}}{x} &= \funElse{x}\label{eq:strong:elseValue}
\end{align}
\end{defi}

Note that \Cref{eq:weak:thenValue} is \Cref{eq:strong:thenValue} and \Cref{eq:weak:elseLadder} is \Cref{eq:strong:elseLadder}.
Without surprise, if $\symbSecond$ is chosen such that $\funSecond{x}{y} = \funElse{y}$ then \Cref{eq:weak:thenLadder} is a special case of \Cref{eq:strong:thenLadder}, and \Cref{eq:strong:elseValue} is satisfied.
Thus, semi-interleaved ladders are subcases of fully-interleaved ladders.

\subsection{Ladderizable Algorithms}
\label{sec:ladderEquations:th1}

\begin{theo}
Let $A$ be an algorithm with an iterative conditional branching with two unary functions denoted $\symbThen$ and $\symbElse$.
If $A$ is semi-ladderizable with $\symbLadder$ and $\symbFirst$, or fully-ladderizable with $\symbLadder$, $\symbFirst$ and $\symbSecond$, then for every iteration of the ladder variant:
\begin{itemize}
    \item $y = \funLadder{x}$
    \item $x$ is updated as in $A$:
        \[x \leftarrow \left\{
            \begin{array}{rl}
                \funThen{x}	&\text{if secret}\\
                \funElse{x}		&\text{otherwise}
            \end{array}
        \right.\]
\end{itemize}
\end{theo}

Thus, for every algorithm with an iterative conditional branching, 
if there exists $\symbLadder$ and $\symbFirst$ satisfying the equations in \Cref{defi:weakLadder} then the conditional branching can be rewritten
as a semi-interleaved ladder. Even better, if there exists $\symbLadder$, $\symbFirst$ and $\symbSecond$ satisfying the equations in \Cref{defi:strongLadder} then it can be rewritten
as a fully-interleaved ladder.

Because in a semi- or fully-interleaved ladder the operations performed are the same for both the \code{then} and the \code{else} branches, this transformation is an algorithmic countermeasure against side-channel attacks \cite{Koc96,KJJ99}.
We detail in \Cref{sec:faultInjection} the impact of the semi- and fully-interleaved ladderization against fault injection techniques.
Then, to demonstrate the concept, we construct examples of semi- and fully-interleaved ladders in \Cref{sec:exponentiation,sec:scalarMult}.

\section{Fault Injection}
\label{sec:faultInjection}

In this section, we introduce two attacker models using fault injection techniques, and compare the vulnerability of the non-, semi- and fully-interleaved ladders.


According to \cite{ZAV04} a \emph{fault} is a physical defect, imperfection or flaw that occurs within some hardware or software component, while an \emph{error} is a deviation from accuracy or correctness, and is the manifestation of a fault.
Hardware/physical faults can be permanent, transient or intermittent, while software faults are the consequence of incorrect design, at speci\-fication or at coding time.
\emph{Fault injection} is defined \cite{Arl90} as the validation technique of the dependability of fault tolerant systems, which consists in performing controlled experiments where the observation of the system’s behavior in presence of faults is induced explicitly by the writing introduction (called injection) of faults in the system.


\subsection{First Attacker Model}
\label{sec:faultInjection:attacker1}

For the iterative conditional branching (\Cref{table:iterCondBranch}) and the semi- and fully-interleaved ladders (\Cref{table:weakLadder,table:strongLadder}), we assume as in the Montgomery ladder (\Cref{table:montgomeryLadder}) that ``$\secret$" is the condition $\bit{k}{i} = 1$, where $\bit{k}{i}$ is the $i$-th bit of the secret key $k$.

\begin{defi}[First Attacker Model]
     \label{defi:FIattack1}
     We assume that:
     \begin{itemize}
         \item The attacker wants to obtain the secret key stored in the chip and copied in the register $k$.
         \item The attacker can run the program any number of times:
         \begin{itemize}
             \item inputting $\xInit$ and $\yInit$, the initial values in the register $x$ and $y$,
             \item obtaining $\xFinal$ and/or $\yFinal$, the final value(s) returned by the program.
         \end{itemize}
         \item A run consists of iterations $i$ over%
         \footnote{To simplify the notations, we assume in this \namecref{sec:faultInjection} that the counter is incremented (from $\iInit$ to $\iFinal$ with a step $1$), but the argument is similar for other initial or final values, a decremented counter, and/or other step values.}
         $\iInit,\dots,\iFinal$, where:
         \begin{itemize}
             \item $\bit{k}{i}$ is the $i$-th bit of $k$,
             \item $x_i$ (resp. $y_i$) denotes the value in the register $x$ (resp. $y$) between the iterations $i-1$ and $i$.
         \end{itemize}
         \item The attacker can $\FI{x_i}$ (resp. $\FI{y_i}$) inject a random fault%
                 \footnote{In the Montgomery ladder (\Cref{table:montgomeryLadder}) the invariant $y = ax$ is satisfied for every iteration. So, if a random fault is injected in $x$ or $y$, the violation of the invariant allows the algorithm to detect the fault \cite{Gir06} and thus to enhance the appropriate fault policy, as stopping the program or switching to a random key for the rest of the computation. The same argument holds for our semi- and fully-interleaved ladders with the invariant $y = \funLadder{x}$.}
                 (the affected variable is set to a random value)
                 in the register of $x$ (resp. $y$) between the iterations $i-1$ and $i$.
     \end{itemize}
\end{defi}

The attacker might be unlucky and obtain the same value as before, but this is unlikely and can be fixed by several tries, so we will assume for simplicity that the new value is different.

The attacker can run a program for the square-and-multiply(-always) (\Cref{table:squareMultiply,table:squareMultiplyAlways}) algorithm(s) for a given input $\xInit$, obtain a value $\xFinal$, then run the program again with the same input while injecting a fault $\FI{y_i}$ in the register $y = ax$ between the iterations $i-1$ and $i$, and obtain a value $\xFault$.
If $\xFault = \xFinal$ then $\bit{k}{i} = 0$, otherwise $\bit{k}{i} = 1$.
This can be done for every iteration (in any order), thus the attacker can obtain that way all the bits of the secret key.

In this algorithm, because the current value in $x$ determines the next value in $x$, a faulted value $\FI{x_i}$ for an iteration $i$ always propagates to the next iteration $\FI{x_{i+1}}$, which we denote by $\FI{x_i} \Rightarrow \FI{x_{i+1}}$.
To obtain the bit $\bit{k}{i}$, the attacker has exploited the fact that a fault in $y = ax$ propagates to $x$, denoted by $\FI{y_i} \Rightarrow \FI{x_{i+1}}$, only if $\bit{k}{i} = 1$.
As opposed to this non-ladderized variant, in \Cref{table:weakLadder,table:strongLadder} the values of $x$ and $y$ are interleaved, and the fault propagation patterns are the following:
\begin{enumerate}
    \item For the semi-interleaved ladder:
         \[\begin{array}{r@{}l}
             \FI{x_i} \Rightarrow \FI{x_{i+1}} &\text{ and } (\FI{y_{i+1}}	 \text{ only if } \bit{k}{i}= 0)\\
             \FI{y_i} \Rightarrow \FI{y_{i+1}} &\text{ and } (\FI{x_{i+1}}	 \text{ only if } \bit{k}{i}= 1)\\
         \end{array}\]
         which means that a fault always propagates to the same register, but propagates to the other depending on the current bit of the secret key.
    \item For the fully-interleaved ladder:
        \[\FI{x_i} \text{ or } \FI{y_i} \Rightarrow \FI{x_{i+1}} \text{ and } \FI{y_{i+1}}\]
        which means that any fault in one register propagates in every case to both.
\end{enumerate}

The fully-interleaved ladder has a lower \emph{fault tolerance}, i.e. it is easier to disrupt the computation.
This is not convenient for properties like functionality, availability or redundancy, but prevents an attacker from obtaining the secret key, thus increases security.
Thus, we reduced the information leakage from fault injection by reducing also the fault tolerance.
The semi-interleaved ladder is more robust, but this comes at the price of a fault propagation pattern depending on the secret key, which can be exploited.

\begin{figure}[htbp]
    \centering
    \includegraphics[width=0.49\linewidth]{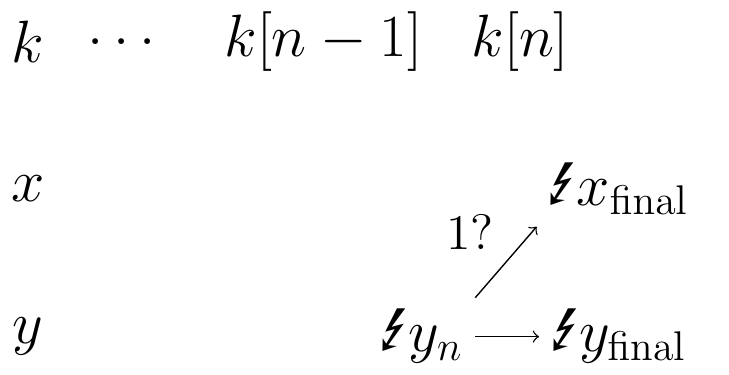}
    \caption{Attack the Last Bit}
    \label{fig:FIweakLast}
\end{figure}

Indeed, the attacker can $\FI{y_{\iFinal}}$ as in \Cref{fig:FIweakLast} and compare the $x$ output.
We assume, as in the exponentiation examples, that changing a read intermediate value would lead to a different output, because if this was not the case then the attacker should have used better $\xInit$ and $\yInit$ inputs.
So, if $\xFault = \xFinal$ then $\bit{k}{\iFinal} = 0$, otherwise $\bit{k}{\iFinal} = 1$.
Thus, the attacker can obtain $\bit{k}{\iFinal}$, the last bit of the secret key.

\begin{figure}[htbp]
    \centering
    \includegraphics[width=0.49\linewidth]{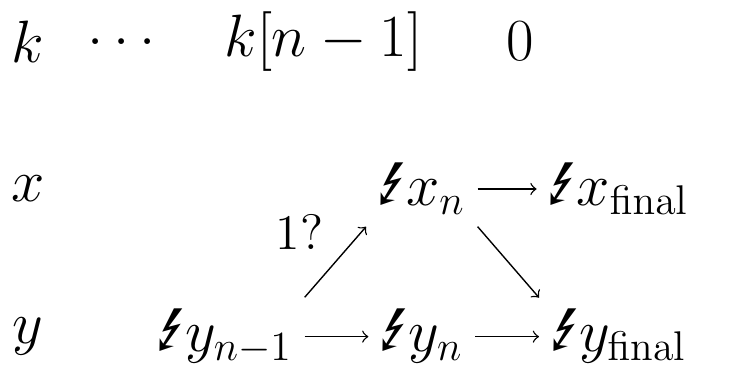}
    \hfill
    \unskip\vrule
    \includegraphics[width=0.49\linewidth]{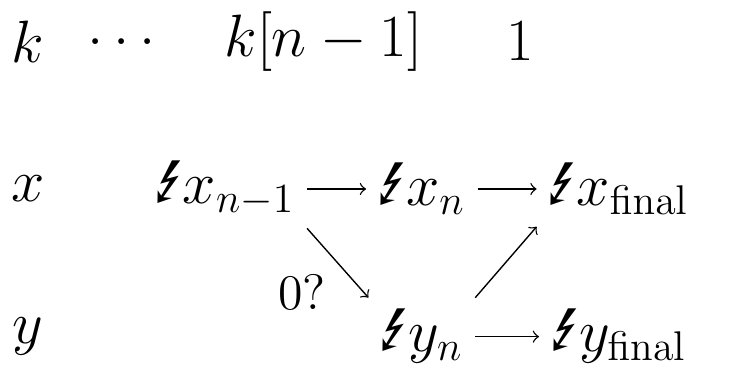}
    \caption{Attack the Penultimate Bit}
    \label{fig:FIweakBeforeLast}
\end{figure}

If the obtained bit was $0$ as in the left part of \Cref{fig:FIweakBeforeLast}, the attacker can $\FI{y_{\iFinal-1}}$.
In that case if $\xFault = \xFinal$ then $\bit{k}{\iFinal-1} = 0$, otherwise $\bit{k}{\iFinal-1} = 1$.
This process can be repeated until a $1$ is found.
If the obtained bit was $1$ as in the right part of \Cref{fig:FIweakBeforeLast}, the attacker can $\FI{x_{\iFinal-1}}$.
In that case if $\yFault = \yFinal$ then $\bit{k}{\iFinal-1} = 1$, otherwise $\bit{k}{\iFinal-1} = 0$.
This process can be repeated until a $0$ is found.

\begin{figure}[htbp]
    \centering
    \includegraphics[width=0.49\linewidth]{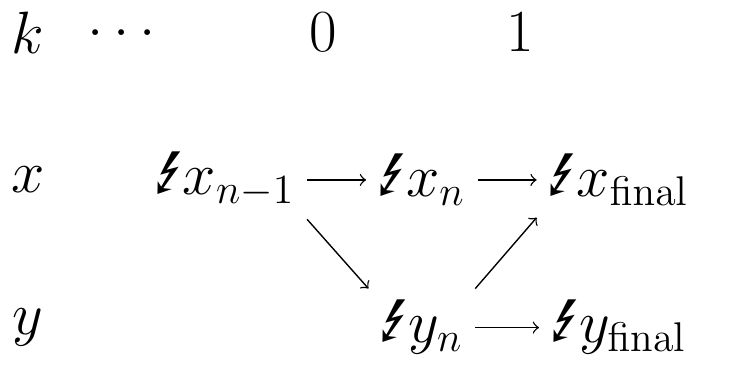}
    \hfill
    \unskip\vrule
    \includegraphics[width=0.49\linewidth]{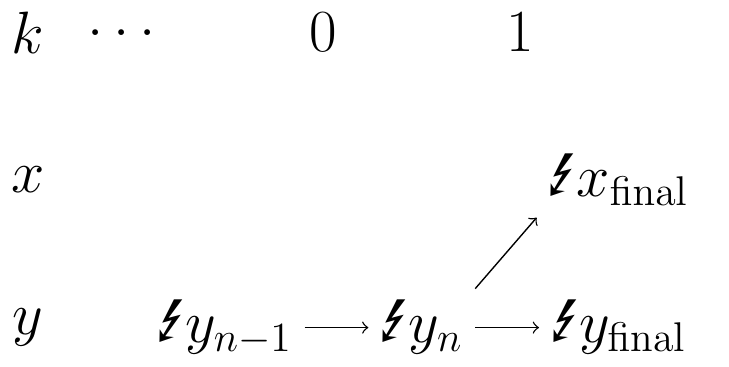}
    \caption{$\FI{x}$ or $\FI{y}$ before $01$}
    \label{fig:FIweak01}
\end{figure}

\begin{figure}[htbp]
    \centering
    \includegraphics[width=0.49\linewidth]{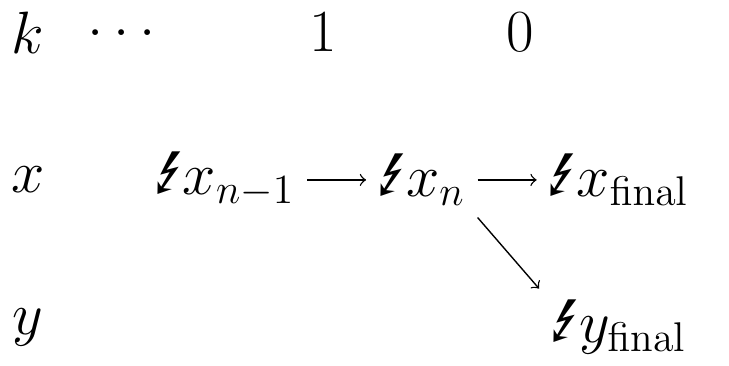}
    \hfill
    \unskip\vrule
    \includegraphics[width=0.49\linewidth]{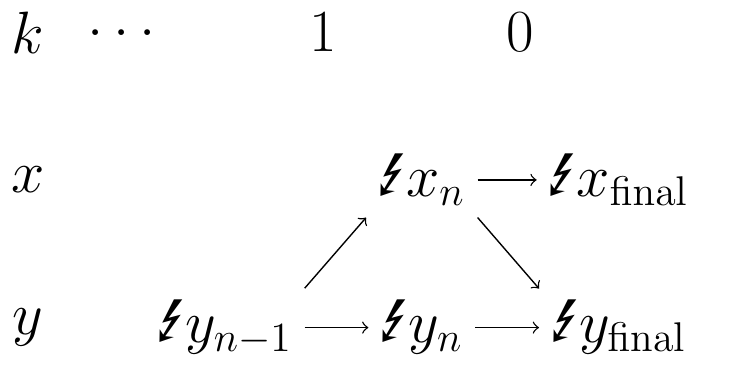}
    \caption{$\FI{x}$ or $\FI{y}$ before $10$}
    \label{fig:FIweak10}
\end{figure}

One may think that these processes can be alternated in order to recover all the bits of the secret key, but if there is a bit alternation, i.e. $\left(\bit{k}{i},\bit{k}{i+1}\right) = (0,1)$ or $(1,0)$, then the bits before $i$ cannot be obtained, as illustrated in \Cref{fig:FIweak01,fig:FIweak10}.
So, the attacker can obtain the final bits $1 0 \dots 0$ if $\xFinal$ can be read, and the final bits $0 1 \dots 1$ if $\yFinal$ can be read.

The vulnerability to fault injection is summarized in \Cref{table:FIvulnAttack1}.
Against the first attacker model described in \Cref{defi:FIattack1}, fully-interleaved ladders are more secure than semi-interleaved ladders, which are more secure than without interleaving at all.
But we show in the next subsection that a stronger attacker is able to obtain all the bits of the key from both the semi- and fully-interleaved ladders.

\subsection{Second Attacker Model}
\label{sec:faultInjection:attacker2}

\begin{defi}[Second Attacker Model]
     \label{defi:FIattack2}
     We assume that:
     \begin{itemize}
         \item The second attacker has the same goal and means that the attacker in \Cref{defi:FIattack1}.
         \item The second attacker can also $\FI{k_{> i}} = 0$ (resp. $\FI{k_{> i}} = 1$) \emph{stuck-at} \cite{ZAV04} $0$ (resp. $1$) all the bits of the register $k$ between iterations $i$ and $i+1$.
     \end{itemize}
\end{defi}

\begin{table}[t]
    \begin{algorithmic}[1]
        \STATE $\text{FIx} \leftarrow 0$
        \FOR{$i = \iFinal$ \TO $\iInit$}
            \IF{$\text{FIx} = 1$}
                \STATE $(\xFinal,\yFinal) \leftarrow \run(\xInit,\yInit,\FI{k_{>i}} = 1)$\label{algo:line:run}
                \STATE $(\xFault,\yFault) \leftarrow \run(\xInit,\yInit,\FI{k_{>i}} = 1,\FI{x_i})$\label{algo:line:runFI}
                \IF{$\yFault = \yFinal$}
                    \STATE $\bit{k}{i} \leftarrow 1$
                \ELSE
                    \STATE $\bit{k}{i} \leftarrow 0$
                    \STATE $\text{FIx} \leftarrow 0$
                \ENDIF
            \ELSE
                \STATE $(\xFinal,\yFinal) \leftarrow \run(\xInit,\yInit,\FI{k_{>i}} = 0)$
                \STATE $(\xFault,\yFault) \leftarrow \run(\xInit,\yInit,\FI{k_{>i}} = 0,\FI{y_i})$
                \IF{$\xFault = \xFinal$}
                    \STATE $\bit{k}{i} \leftarrow 0$
                \ELSE
                    \STATE $\bit{k}{i} \leftarrow 1$
                    \STATE $\text{FIx} \leftarrow 1$
                \ENDIF
            \ENDIF
        \ENDFOR
        \RETURN $k$
    \end{algorithmic}
    \caption{Protocol to Attack the Semi-Interleaved Ladders}
    \label{table:attackWeak}
\end{table}

This stronger attacker is able to break the semi-interleaved ladders by using the attack protocol described in \Cref{table:attackWeak} in order to recover all the bits of the secret key%
\footnote{Note that the iterations are reversed in this attack protocol.}. 
$\run(\xInit,\yInit,\FI{k_{>i}} = 1)$ at Line~\ref{algo:line:run} means that the studied program is executed with inputs $\xInit,\yInit$ and a stuck-at $\FI{k_{>i}} = 1$, and $\run(\xInit,\yInit,\FI{k_{>i}} = 1,\FI{x_i})$ at Line~\ref{algo:line:runFI} is the same but with a fault $\FI{x_i}$.
In particular, if both $\xFinal$ and $\yFinal$ can be read, then the Montgomery Ladder (\Cref{table:montgomeryLadder}) can be broken (by iterating over $0$ to $d$).
This attack does not work against fully-interleaved ladders, but a fully-interleaved ladder is more difficult to obtain (when possible), as discussed in \Cref{sec:exponentiation:strongCase,sec:scalarMult:strongCase}.

%
%

Actually, being able to write deterministically in the key register is enough to obtain with a third attack all the bits of the key, even against fully-interleaved ladders.
Indeed, let's assume that the second attacker knows the first $i$ bits $\bit{k}{1} \dots \bit{k}{i}$, where $i = 0$ for the initialization.
By $\FI{k_{> i}} = 0$ (resp. $\FI{k_{> i}} = 1$) and $\FI{k_{> i+1}} = 0$ (resp. $\FI{k_{> i+1}} = 1$) the second attacker can compare ${\xFault}_0$ and ${\xFinal}_0$ (resp. ${\xFault}_1$ and ${\xFinal}_1$) obtained respectively for $\bit{k}{1} \dots \bit{k}{i} 0\ 0 \dots 0$ and $\bit{k}{1} \dots \bit{k}{i} \bit{k}{i+1} 0 \dots 0$ (resp. with $1$ instead of $0$ at the end), and the same for $y$.
Thus, the second attacker can (almost) determine whether $\bit{k}{i+1} = 1$ (resp. $0$):
\begin{itemize}
    \item if ${\xFault}_0 \not= {\xFinal}_0$ or ${\yFault}_0 \not= {\yFinal}_0$ then $\bit{k}{i+1} = 1$,
    \item if ${\xFault}_1 \not= {\xFinal}_1$ or ${\yFault}_1 \not= {\yFinal}_1$ then $\bit{k}{i+1} = 0$.
\end{itemize}
If ${\xFault}_0 = {\xFinal}_0$, ${\xFault}_1 = {\xFinal}_1$, ${\yFault}_0 = {\yFinal}_0$ and ${\yFault}_1 = {\yFinal}_1$ then another inputs $\xInit,\yInit$ should be chosen in the protocol described in \Cref{table:attackStrong}.

\begin{table}[t]
    \begin{algorithmic}[1]
        \FOR{$i = 0$ \TO $\iFinal - 1$}
            \STATE $(\xFinal,\yFinal) \leftarrow \run(\xInit,\yInit,\FI{k_{>i}} = 0)$
            \STATE $(\xFault,\yFault) \leftarrow \run(\xInit,\yInit,\FI{k_{>{i+1}}} = 0)$    
            \IF{$\xFault \not= \xFinal$ or $\yFault \not= \yFinal$}
                \STATE $\bit{k}{i+1} \leftarrow 1$
            \ELSE
                \STATE $(\xFinal,\yFinal) \leftarrow \run(\xInit,\yInit,\FI{k_{>i}} = 1)$
                \STATE $(\xFault,\yFault) \leftarrow \run(\xInit,\yInit,\FI{k_{>{i+1}}} = 1)$    
                \IF{$\xFault \not= \xFinal$ or $\yFault \not= \yFinal$}
                    \STATE $\bit{k}{i+1} \leftarrow 0$
                \ENDIF
            \ENDIF
        \ENDFOR
        \RETURN $k$
    \end{algorithmic}
    \caption{Protocol to Attack Both Semi- and Fully-Interleaved Ladders}
    \label{table:attackStrong}
\end{table}

That way all the bits of the key $k$ can be obtained, except the ones such that for every possible $\xInit$, ${\xFault}_0 = {\xFinal}_0$ and ${\xFault}_1 = {\xFinal}_1$.
But this case is unlikely, it does not prevent the attacker from discovering the other bits, and these bits make no difference in the result, thus probably the attacker does not care.
Therefore, the second attacker can obtain all the (relevant) bits of the key, even against semi- and fully-interleaved ladders.
The vulnerabilities against these attacks are summarized in \Cref{table:FIvulnAttack1}.

\begin{table}[t]
    \centering
    \begin{tabular}{c|c|c|c|}
        \cline{2-4}
            &\multicolumn{3}{c|}{Obtained Bits}\\
        \cline{2-4}
            &Model 1 &\multicolumn{2}{c|}{Model 2}\\
        \hline
        \multicolumn{1}{|c|}{Interleaving}
            &Attack 1 &Attack 2 &Attack 3\\
        \hline
        \multicolumn{1}{|c|}{non}
            &all & all &all\\
        \multicolumn{1}{|c|}{semi}
            &some &all &all\\
        \multicolumn{1}{|c|}{fully}
            &none &none &all\\
        \hline
    \end{tabular}
    \vspace{0.2cm}
    \caption{Vulnerability against Fault-Injection Attacks}
    \label{table:FIvulnAttack1}
\end{table}

\section{Extending the Montgomery Ladder}
\label{sec:exponentiation}

The purpose of this section is to provide concrete examples of the ladder equations for cryptography, and to generalize the idea behind the Montgomery ladder to improve its protection against side-channel and fault injection attacks.

We demonstrate that by using the modular exponentiation.
In the following for any $n$ we consider $\ZnZ{n}$, i.e. integers modulo $n$, and denote $\ZnZinv{n}$ the integers invertible modulo $n$.
We assume that $\funThen{x}$, $\funElse{x}$, $\funFirst{x}{y}$, $\funSecond{x}{y}$ and $\funLadder{x}$ are quadratic polynomials with the following coefficients:
\begin{align}
    \funThen{x} &= \cThenA x^2\nonumber\\
    \funElse{x} &= x^2\nonumber\\
    \funFirst{x}{y} &= \cFirstA x^2 + \cFirstB xy + \cFirstC y^2 + \cFirstD x + \cFirstE y + \cFirstF\nonumber\\
    \funSecond{x}{y} &= \cSecondA x^2 + \cSecondB xy + \cSecondC y^2 + \cSecondD x + \cSecondE y + \cSecondF\nonumber\\
    \funLadder{x} &= \cLadderB x + \cLadderC\nonumber
\end{align}

More general quadratic polynomials, e.g. $\cLadderA \not= 0$ or more complex $\funThen{x}$ and $\funElse{x}$, can be investigated
but the general systems of equations tend to be complicated, and in this \namecref{sec:exponentiation} we want to focus on the exponentiation algorithms.
To ensure that $\funThen{x} = \cThenA x^2$ and $\funLadder{x} = \cLadderB x + \cLadderC$ depends on $x$, we will assume that $\cThenA \not= 0$ and $\cLadderB \not= 0$.
Moreover, in the following we focus on solutions without constraint on $\cThenA$ to preserve the generality of the original algorithm.

Finally, note that $\ZnZ{n}$ may not be an integral domain i.e. there exists $x, y \in \ZnZ{n} \setminus \set{0}{}$ such that $xy = 0 \bmod n$, which is the case for RSA (see \Cref{sec:exponentiation:RSA}) with $p$ and $q$ such that $n = pq$.
Because $n$ cannot be easily factorized into $p$ and $q$, and because we are looking for general solutions for any $n$, if $x \not= 0$ and $xy = 0$ we will look for solutions $y = 0$, thus loosing some generality but simplifying the analysis.

\subsection{Both Semi- and Fully-Interleaved Ladders}
\label{sec:exponentiation:commonEqs}

%
\Cref{eq:weak:thenValue} (resp. \Cref{eq:strong:thenValue}) $\funFirst{x}{\funLadder{x}} = \funThen{x}$ for the semi- (resp. fully-) interleaved cases is equivalent to:
\begin{align}[left=\empheqlbrace]
    \cFirstA + \cFirstB\cLadderB + \cFirstC\cLadderB^2
        &= \cThenA\nonumber\\
    \cFirstB\cLadderC + 2\cFirstC\cLadderB\cLadderC+ \cFirstD + \cFirstE\cLadderB
        &= 0\nonumber\\
    \cFirstC\cLadderC^2 + \cFirstE\cLadderC + \cFirstF
        &=0\nonumber
\end{align}
%
\Cref{eq:weak:elseLadder} (resp. \Cref{eq:strong:elseLadder}) $\funFirst{\funLadder{x}}{x} = \funLadder{\funElse{x}}$ for the  semi- (resp. fully-) interleaved cases is equivalent to:
\begin{align}[left=\empheqlbrace]
    \cFirstA\cLadderB^2 + \cFirstB\cLadderB + \cFirstC
        &= \cLadderB\nonumber\\
    2\cFirstA\cLadderB\cLadderC + \cFirstB\cLadderC + \cFirstD\cLadderB + \cFirstE
        &= 0\nonumber\\
    \cFirstA\cLadderC^2 + \cFirstD\cLadderC + \cFirstF
        &=\cLadderC\nonumber
\end{align}

\subsection{Semi-Interleaved Ladder}
\label{sec:exponentiation:weakCase}

%
The remaining \Cref{eq:weak:thenLadder} $\funElse{\funLadder{x}} = \funLadder{\funThen{x}}$ for the semi-interleaved cases is equivalent to:
\begin{align}[left=\empheqlbrace]
    \cLadderB^2
        &= \cLadderB\cThenA\nonumber\\
    2\cLadderB\cLadderC
        &= 0\nonumber\\
    \cLadderC^2
        &=\cLadderC\nonumber
\end{align}

So, because $\cLadderB \not= 0$, we assume $\cLadderB = \cThenA$ and $\cLadderC = 0$.
Therefore \Cref{eq:weak:thenValue} is equivalent to:
\begin{align}[left=\empheqlbrace]
    \cFirstA + \cFirstB\cThenA + \cFirstC\cThenA^2
        &= \cThenA\nonumber\\
    \cFirstD + \cFirstE\cThenA
        &= 0\nonumber\\
    \cFirstF
        &=0\nonumber
\end{align}
and \Cref{eq:weak:elseLadder} is equivalent to:
\begin{align}[left=\empheqlbrace]
    \cFirstA\cThenA^2 + \cFirstB\cThenA + \cFirstC
        &= \cThenA\nonumber\\
    \cFirstD\cThenA + \cFirstE
        &= 0\nonumber\\
    \cFirstF
        &=0\nonumber
\end{align}

From $\cFirstD + \cFirstE\cThenA = \cFirstD\cThenA + \cFirstE$ we deduce $\cFirstD(\cThenA - 1) = \cFirstE(\cThenA - 1)$, which is always true without constraint on $\cThenA$ if $\cFirstD = \cFirstE$.
Moreover, from $\cFirstD + \cFirstE\cThenA = 0$ we obtain $\cFirstD = - \cFirstE\cThenA$, thus from $\cFirstD\cThenA + \cFirstE = 0$ we obtain $\cFirstE(\cThenA^2 - 1) = 0$, which is always true without constraint on $\cThenA$ if $\cFirstE = 0$.
Therefore we assume $\cFirstD = 0 = \cFirstE$.

From $\cFirstA + \cFirstB\cThenA + \cFirstC\cThenA^2 = \cFirstA\cThenA^2 + \cFirstB\cThenA + \cFirstC$ we deduce $\cFirstA(\cThenA^2 - 1) = \cFirstC(\cThenA^2 - 1)$, which is always true without constraint on $\cThenA$ if $\cFirstA = \cFirstC$.
The remaining constraint is $\cFirstA(\cThenA^2 + 1) + \cFirstB\cThenA = \cThenA$, which is equivalent to $\cFirstA = \frac{\cThenA}{\cThenA^2 + 1}(1 - \cFirstB)$.
By assuming that the coefficients are integers there exists $\cWeakB$ such that $1 - \cFirstB = \cWeakB(\cThenA^2 + 1)$. Thus, we have $\cFirstA = \cWeakB\cThenA$ and $\cFirstB = 1 - \cWeakB(\cThenA^2 + 1)$.

\begin{theo}
\label{theo:expSIL}
The square-and-multiply algorithm in \Cref{table:squareMultiplyAlways} for the modular exponentiation with $\funThen{x} = \cThenA x^2$ and $\funElse{x} = x^2$ is semi-ladderizable with:
    \begin{align}
        \funLadder{x} &= \cThenA x\nonumber\\
        \funFirst{x}{y} &= \cWeakB\cThenA (x^2 + y^2) + (1 - \cWeakB(\cThenA^2 + 1)) xy\nonumber
    \end{align}
producing the algorithm in \Cref{table:weakExample}.
\end{theo}

Note that if $\cWeakB = 0$ then $\funFirst{x}{y} = xy$ so this solution is the common Montgomery ladder. 
Note also that $\cWeakB$ can be chosen randomly for every iteration, providing a mask for the intermediate va\-lues and thus reducing the opportunity of information leakage.


\begin{table}[t]
    \begin{algorithmic}[1]
        \REQUIRE $\public a, n; \secret k$
        \STATE $x \leftarrow 1$
        \STATE $y \leftarrow a \bmod n$
        \STATE $c \leftarrow a^2 + 1 \bmod n$
        \FOR{$i = d$ \TO $0$}
            \STATE $\cWeakB \leftarrow \rand([0, n - 1])$
            \IF{$\bit{k}{i} = 1$}
                \STATE $z \leftarrow y^2 \bmod n$
                \STATE $x \leftarrow \cWeakB a(x^2 + z) + (1 - \cWeakB c)xy \bmod n$
                \STATE $y \leftarrow z$
            \ELSE
                \STATE $z \leftarrow x^2 \bmod n$
                \STATE $y \leftarrow \cWeakB a(y^2 + z) + (1 - \cWeakB c)yx \bmod n$
                \STATE $x \leftarrow z$
            \ENDIF
        \ENDFOR
        \RETURN $x$
        \ENSURE $x = a^k \bmod n$
    \end{algorithmic}
    \caption{Semi-Interleaved Ladder for the Exponentiation}
    \label{table:weakExample}
\end{table}

\subsection{Fully-Interleaved Ladder (General Case)}
\label{sec:exponentiation:strongCase}

\Cref{eq:strong:thenLadder} $\funSecond{\funThen{x}}{\funLadder{x}} = \funLadder{\funThen{x}}$ for the fully-interleaved cases is equivalent to:
\begin{align}[left=\empheqlbrace]
    \cSecondA\cThenA^2
        &= 0\nonumber\\
    \cSecondB\cLadderB\cThenA
        &= 0\nonumber\\
    \cSecondB\cLadderC\cThenA + \cSecondC\cLadderB^2 + \cSecondD \cThenA
        &=\cLadderB\cThenA\nonumber\\
    2\cSecondC\cLadderB\cLadderC + \cSecondE\cLadderB
        &=0\nonumber\\
    \cSecondC\cLadderC^2 + \cSecondE\cLadderC + \cSecondF
        &=\cLadderC\nonumber
\end{align}
So, because $\cThenA \not= 0$ and $\cLadderB \not= 0$, we assume $\cSecondA = 0$ from the first equation, $\cSecondB = 0$ from the second, and the remaining constraints are:
\begin{align}[left=\empheqlbrace]
    \cSecondC\cLadderB^2 + \cSecondD \cThenA
        &=\cLadderB\cThenA\nonumber\\
    2\cSecondC\cLadderB\cLadderC + \cSecondE\cLadderB
        &=0\nonumber\\
    \cSecondC\cLadderC^2 + \cSecondE\cLadderC + \cSecondF
        &=\cLadderC\nonumber
\end{align}

According to \Cref{eq:strong:elseLadder} $\funFirst{\funLadder{x}}{x} = \funLadder{\funElse{x}} =  \cLadderB x^2 + \cLadderC$, so by using $\cSecondA = \cSecondB = 0$ we have:
\begin{align}
    \funSecond{\funFirst{\funLadder{x}}{x}}{x} =\ 
        &(\cSecondC + \cSecondD\cLadderB)x^2 + (\cSecondE)x + (\cSecondD\cLadderC + \cSecondF)\nonumber
\end{align}
Thus, \Cref{eq:strong:elseValue} $\funSecond{\funFirst{\funLadder{x}}{x}}{x} = \funElse{x}$ for the fully-interleaved cases is equivalent to:
\begin{align}[left=\empheqlbrace]
    \cSecondC + \cSecondD\cLadderB
        &=1\nonumber\\
    \cSecondE
        &=0\nonumber\\
    \cSecondD\cLadderC + \cSecondF
        &=0\nonumber
\end{align}
Because $\cSecondA = \cSecondB = \cSecondE = 0$ and we require $\funSecond{x}{y} = \cSecondC y^2 + \cSecondD x + \cSecondF$ to depend on both $x$ and $y$, we have to assume $\cSecondC \not= 0$ and $\cSecondD \not= 0$.
By using $\cSecondE = 0$ the remaining constraints from \Cref{eq:strong:thenLadder} are:
\begin{align}[left=\empheqlbrace]
    \cSecondC\cLadderB^2 + \cSecondD\cThenA
        &=\cLadderB\cThenA\nonumber\\
    \cSecondC\cLadderB\cLadderC
        &=0\nonumber\\
    \cSecondC\cLadderC^2 + \cSecondF
        &=\cLadderC\nonumber
\end{align}
Because $\cSecondC \not= 0$ and $\cLadderB \not= 0$, the second equation implies that $\cLadderC = 0$. Thus, the remaining constraints from \Cref{eq:strong:thenLadder} and \Cref{eq:strong:elseValue} are:
\begin{align}[left=\empheqlbrace]
    \cSecondC\cLadderB^2 + \cSecondD\cThenA
        &=\cLadderB\cThenA\nonumber\\
    \cSecondC + \cSecondD\cLadderB
        &=1\nonumber\\
    \cSecondA = \cSecondB = \cSecondE = \cSecondF
        &=0\nonumber
\end{align}
By using the second equation $\cSecondC = 1 - \cSecondD\cLadderB$, the first one is equivalent to:
\begin{align}
    \cSecondD(\cThenA - \cLadderB^3)
        &= \cLadderB(\cThenA - \cLadderB)\nonumber
\end{align}
Because $\cSecondD \not= 0$ and $\cLadderB \not= 0$, note that $\cThenA = \cLadderB^3 \Leftrightarrow \cThenA = \cLadderB$, in which case $\cLadderB^3 = \cLadderB$, thus $\cThenA = \cLadderB = \pm 1$.
To remove the constraint on $\cThenA$, we assume that $\cThenA \not= \cLadderB^3$, and thus $\cSecondD$ and $\cSecondC$ depends only on $\cThenA$ and $\cLadderB$:
\begin{align}
    \cSecondD
        &= \cLadderB\frac{\cThenA - \cLadderB}{\cThenA - \cLadderB^3}\nonumber\\
    \cSecondC
        &= 1 - \cSecondD\cLadderB = \cThenA\frac{1 - \cLadderB^2}{\cThenA -  \cLadderB^3}\nonumber\\
    \funSecond{x}{y}
        &= \cSecondC y^2 + \cSecondD x\nonumber\\
        &= \frac{1}{\cThenA -  \cLadderB^3}\left(\cThenA(1 - \cLadderB^2) y^2 + \cLadderB(\cThenA - \cLadderB)x\right)\nonumber
\end{align}
Note that because $\cLadderB \not= 0$ and $\cSecondD \not= 0$ we have to assume that, as opposed to the semi-interleaved cases, $\cLadderB \not= \cThenA$.
Moreover, because $\cLadderC = 0$, the constraints from the common \Cref{eq:strong:thenValue,eq:strong:elseLadder} can be simplified:
\begin{align}[left=\empheqlbrace]
    \cFirstA + \cFirstB\cLadderB + \cFirstC\cLadderB^2
        &= \cThenA\tag{$L_1$}\\
    \cFirstA\cLadderB^2 + \cFirstB\cLadderB + \cFirstC
        &= \cLadderB\tag{$L_2$}\\
     \cFirstD + \cFirstE\cLadderB
        &= 0\tag{$L_3$}\\
    \cFirstD\cLadderB + \cFirstE
        &= 0\tag{$L_4$}\\
    \cFirstF
        &=0\nonumber
\end{align}

In order to have a non-trivial ladder $\funLadder{x} = \cLadderB x$, we will assume that $\cLadderB \not= \pm 1$. With $(L_4) - (L_3)$ we obtain $(\cFirstD - \cFirstE)(\cLadderB - 1) = 0$ thus $\cFirstD = \cFirstE$, and with $(L_3)$ we have $\cFirstD(\cLadderB + 1) = 0$, so $\cFirstD = \cFirstE = 0$. With $(L_1) - (L_2)$ we obtain $(\cFirstC - \cFirstA)(\cLadderB^2 - 1) = \cThenA - \cLadderB$, so $\cFirstC = \cFirstA + \frac{\cThenA - \cLadderB}{\cLadderB^2 - 1}$. Therefore, remain the following constraints:
\begin{align}[left=\empheqlbrace]
    \cFirstA(\cLadderB^2 + 1) + \cFirstB\cLadderB
        &= \cThenA - \cLadderB^2\frac{\cThenA - \cLadderB}{\cLadderB^2 - 1}\nonumber\\
        &= \cLadderB - \frac{\cThenA - \cLadderB}{\cLadderB^2 - 1}\nonumber\\
    \cFirstC
        &= \cFirstA + \frac{\cThenA - \cLadderB}{\cLadderB^2 - 1}\nonumber\\
    \cFirstD = \cFirstE = \cFirstF
        &=0\nonumber
\end{align}
The first line being obtained from $(L_1)$ and the second from $(L_2)$.
Note that $\cThenA - \cLadderB^2\frac{\cThenA - \cLadderB}{\cLadderB^2 - 1} = \cLadderB - \frac{\cThenA - \cLadderB}{\cLadderB^2 - 1}$ is satisfied and thus is not a constraint.
Moreover, because $\cLadderB \not= 0$, by using the second line, $\cFirstB$ can be written as:
\begin{align}
     \cFirstB
        &= \frac{1}{\cLadderB}\left(\cLadderB - \frac{\cThenA - \cLadderB}{\cLadderB^2 - 1} - \cFirstA(\cLadderB^2 + 1)\right)\nonumber\\
        &= \frac{1}{\cLadderB}\left(\frac{\cLadderB^3 - \cThenA}{\cLadderB^2 - 1} - \cFirstA(\cLadderB^2 + 1)\right)\nonumber
\end{align}
Thus, we have:
\begin{align}
    \funFirst{x}{y}%
         ={}&\cFirstA x^2 + \cFirstB xy + \cFirstC y^2\nonumber\\
         ={}&\cFirstA x^2 + \frac{1}{\cLadderB}\left(\frac{\cLadderB^3 - \cThenA}{\cLadderB^2 - 1} - \cFirstA(\cLadderB^2 + 1)\right) xy\nonumber\\
         &+ \left(\cFirstA + \frac{\cThenA - \cLadderB}{\cLadderB^2 - 1}\right) y^2\nonumber\\
         ={}&\cFirstA \left(x^2 - \frac{\cLadderB^2 + 1}{\cLadderB}xy + y^2\right)\nonumber\\
         &+ \frac{1}{\cLadderB^2 - 1}\left(\frac{\cLadderB^3 - \cThenA}{\cLadderB}xy + (\cThenA - \cLadderB)y^2\right)\nonumber
\end{align}
Finally, $\cFirstA$ has no constraint.
It could have been used as a mask for $\symbFirst$ as in \Cref{theo:expSIL}, but unfortunately not for $\symbSecond$.
So, for sake of simplicity, we assume $\cFirstA = 0$ and obtain the following functions:

\begin{theo}
\label{theo:expFIL}
The square-and-multiply algorithm in \Cref{table:squareMultiplyAlways} for the modular exponentiation with $\funThen{x} = \cThenA x^2$ and $\funElse{x} = x^2$ is fully-ladderizable with:
\begin{align}
    \funLadder{x} &= \cLadderB x\nonumber\\
    \funFirst{x}{y} &= \frac{1}{\cLadderB^2 - 1}\left(\frac{\cLadderB^3 - \cThenA}{\cLadderB}xy - (\cLadderB - \cThenA)y^2\right)\nonumber\\
    \funSecond{x}{y} &= \frac{1}{\cLadderB^3 - \cThenA}\left(\cThenA(\cLadderB^2 - 1) y^2 + \cLadderB(\cLadderB - \cThenA)x\right)\nonumber
\end{align}
where \condN{1} $\cLadderB \not= \cThenA \bmod n$, \condN{2} $\cLadderB \in \ZnZinv{n}$, \condN{3} $\cLadderB^2 - 1 \in \ZnZinv{n}$ and \condN{4} $\cLadderB^3 - \cThenA \in \ZnZinv{n}$,
producing the algorithm in \Cref{table:strongExample}.
\end{theo}

The first constraint on the ladder constant (denoted $\symbLadder$ in the following) can be satisfied by checking whether $(\symbLadder - \cThenA) \bmod n = 0$.
The other constraints can be satisfied by using the Extended Euclidean Algorithm $(\cGcd,\cInverse) \leftarrow \egcd(\cVal,n)$, where $\cInverse\cVal = \cGcd \bmod n$, such that if $\cGcd = 1$ then $\cVal$ is invertible modulo $n$ and $\cVal^{-1} = \cInverse \bmod n$.
Because $\symbLadder \not= \cThenA \bmod n$,
$0 \not\in \ZnZinv{n}$,
and if $\symbLadder = \pm 1$ then $\cLadderB^2 - 1 \not\in \ZnZinv{n}$,
we can assume that $\symbLadder \in \ZnZ{n}$ can be chosen in $\intervalLadder{a}{n} = [2, n - 2] \setminus \set{a}{}$, then check whether $\symbLadder$ satisfies the constraints of \Cref{theo:expFIL} (Lines~\ref{algo:line:ladderConstant-begin}--\ref{algo:line:ladderConstant-end}).
Actually, we prove in the RSA case (\Cref{sec:exponentiation:RSA}) and the DSA case (\Cref{sec:exponentiation:DSA}) that a ladder constant verifying these constraints is almost always obtained after one iteration.

Note that after a suitable $\symbLadder$ is chosen, the coefficients of $\symbFirst$ and $\symbSecond$ are constant during the iterations, thus can be pre-computed (Lines~\ref{algo:line:coeffs-begin}--\ref{algo:line:coeffs-end}).
These computations to find suitable constants may cause a significant overhead (Lines~\ref{algo:line:overhead-begin}--\ref{algo:line:overhead-end}) but that does not depend on the secret, thus trading execution time and energy consumption for more security.

\begin{table}[t]
    \begin{algorithmic}[1]
        \REQUIRE $\public a, n; \secret k$
        \REPEAT\label{algo:line:overhead-begin}\label{algo:line:ladderConstant-begin}
            \STATE $\symbLadder \leftarrow \rand([2, n - 2] \setminus \set{a}{})$
            \STATE $\cValA \leftarrow \symbLadder - a \bmod n$
            \STATE $(\cGcdB,\cInverseB) \leftarrow \egcd(\symbLadder,n)$
            \STATE $\cValC \leftarrow \symbLadder^2 - 1 \bmod n$
            \STATE $(\cGcdC,\cInverseC) \leftarrow \egcd(\cValC,n)$
            \STATE $\cValD \leftarrow \symbLadder^3 - a \bmod n$
            \STATE $(\cGcdD,\cInverseD) \leftarrow \egcd(\cValD,n)$
        \UNTIL{$\cValA \bmod n = 0 \lor \cGcdB \not= 1 \lor \cGcdC \not= 1 \lor \cGcdD \not= 1$}\label{algo:line:ladderConstant-end}
        \STATE $\cStrongA \leftarrow \cInverseB \cInverseC \cValD \bmod n$\label{algo:line:coeffs-begin}
        \STATE $\cStrongB \leftarrow - \cValA \cInverseC \bmod n$
        \STATE $\cStrongC \leftarrow a \cValC \cInverseD \bmod n$
        \STATE $\cStrongD \leftarrow \symbLadder \cValA \cInverseD \bmod n$\label{algo:line:coeffs-end}
        \STATE $x \leftarrow 1$
        \STATE $y \leftarrow \symbLadder$\label{algo:line:overhead-end}
        \FOR{$i = d$ \TO $0$}
            \IF{$\bit{k}{i} = 1$}
                \STATE $x \leftarrow \cStrongA xy + \cStrongB y^2 \bmod n$\label{algo:line:memoizationA}
                \STATE $y \leftarrow \cStrongC y^2 + \cStrongD x \bmod n$\label{algo:line:memoizationB}
            \ELSE
                \STATE $y \leftarrow \cStrongA yx + \cStrongB x^2 \bmod n$
                \STATE $x \leftarrow \cStrongC x^2 + \cStrongD y \bmod n$
            \ENDIF
        \ENDFOR
        \RETURN $x$
        \ENSURE $x = a^k \bmod n$
    \end{algorithmic}
    \caption{Fully-Interleaved Ladder for the Modular Exponentiation (General Case)}
    \label{table:strongExample}
\end{table}

We provide in \Cref{table:complexity} a comparison of complexities for the Montgomery, semi- and fully-interleaved ladders.
The complexities are given in terms of cost per key bit, where $M$ stands for multiplication, $S$ for squaring and $A$ for addition/subtraction, all modulo $n$.
These costs are given after memoization, for instance the squarings in Line~\ref{algo:line:memoizationA} and Line~\ref{algo:line:memoizationB} in \Cref{table:strongExample} are the same, so could have been memorized by using $z \leftarrow y^2$ and then $z$ used instead of the squares.
Because the costs are given per key bit, we did not include the cost of the pre-computations, which is far from negligible for the fully-interleaved ladder in \Cref{table:strongExample}. 

\begin{table}
    \centering
    \begin{tabular}{|c|c|c|}
        \hline
        Montgomery & Semi-Interleaved & Fully-Interleaved \\
        \hline
        $M + S$ & $5M + 2S + 3A$ & $5M + S + 2A$\\
        \hline
    \end{tabular}
    \vspace{0.2cm}
    \caption{Cost per Bit of the Montgomery, Semi- and Fully-Interleaved Ladders (After Memoization)}
    \label{table:complexity}
\end{table}

%

\subsection{Fully-Interleaved Ladder (RSA Case)}
\label{sec:exponentiation:RSA}

In the context of RSA~\cite{RSA78}, we assume that $n = pq$, where $p,q$ are distinct primes.
The modular exponentiation aims at computing $a^k \bmod n$, so we assume that the input $a$ is already given modulo $n$, i.e. $0 \le a \le n - 1$.
Because $0^k$ and $(\pm 1)^k$ are fairly trivial to compute, to simplify the following reasoning we assume that $2 \le a \le n - 2$, so $\card{\intervalLadder{a}{n}} = n - 4$, where $\intervalLadder{a}{n} = [2, n - 2] \setminus \set{a}{}$.

To satisfy constraints \condN{1} and \condN{2} of \Cref{theo:expFIL}, i.e. $\symbLadder \not= \cThenA$ and $\symbLadder \in \ZnZinv{n}$, $\symbLadder$ has to be chosen in $\intervalLadder{a}{n}$ such that $\pgcd{\symbLadder}{n} = 1$.
Thus, in the following we will assume that $\symbLadder$ is chosen randomly in $\intervalLadder{a}{n}$ and determine the probability that $\symbLadder$ satisfies the ladder constraints.
Within this range, the only integers such that $\pgcd{\symbLadder}{n} > 1$ are $jp$ for $j \in [1, q - 1]$ or $jq$ for $ j \in [1, p - 1]$, so the probability to pick randomly a ladder constant not verifying \condN{2} is:
\begin{align}
    \proba{}{\text{not } \condN{2}}
        &\le \frac{(p - 1) + (q - 1)}{n - 4} \approx \frac{1}{p} + \frac{1}{q}\nonumber
\end{align}

In RSA context, it means that a random $\symbLadder \in \intervalLadder{a}{n}$ is invertible mo\-du\-lo $n$ with probability almost equal to $1$.
Indeed, according to the National Institute of Standards and Technology (NIST) recommendations \cite{BCR+19},
if $\nBits = \floor{\log_2 n} + 1$ denotes the number of bits of $n = pq$, then $p$ and $q$ should verify:
$$\begin{array}{c}
    2^{\frac{\nBits - 1}{2}} < p < 2^{\frac{\nBits}{2}}\\
    2^{\frac{\nBits - 1}{2}} < q < 2^{\frac{\nBits}{2}}\\
\end{array}$$

Note that if one can compute $\symbLadder$ such that $\pgcd{\symbLadder}{n} > 1$ then $\pgcd{\symbLadder}{n} = p$ or $q$, and is able to factorize $n$.
The same problem occurs in RSA cryptosystem: if Alice can generate a message $m$ such that $\pgcd{m}{n} > 1$, where $n$ is Bob's public key, then Alice can factorize $n$ and find Bob's secret key.
Such an event could occur with a probability nearly equal to 0.

To satisfy constraint \condN{3} of \Cref{theo:expFIL}, i.e. $\symbLadder^2 - 1 \in \ZnZinv{n}$, note that according to the chinese remainder theorem the integers $\symbLadder$ such that $\symbLadder^2 - 1$ is not invertible modulo $n$ verify $\symbLadder^2 = 1 \bmod p$ and $\symbLadder^2 = 1 \bmod q$.
There are at most four such integers: $1$, $n - 1$, $\gamma$, $n - \gamma$, where $\gamma = 1 \bmod p$ and $\gamma = -1 \bmod q$.
So, the probability to pick randomly a ladder constant in $\intervalLadder{a}{n}$ not verifying \condN{3} is:
\begin{align}
    \proba{}{\text{not } \condN{3}}
        &\le \frac{2}{n - 4}\nonumber
\end{align}
Finally, to satisfy the last constraint \condN{4} of \Cref{theo:expFIL}, i.e. $\symbLadder^3 - \cThenA \in \ZnZinv{n}$, we use the following \namecref{theo:Gauss} from Gauss:

\begin{defi}[$r^\text{th}$ Residues]
 \label{defi:residues}
 Let $n$ and $r \ge 2$ be two integers.
 An integer $a \in \ZnZinv{n}$ is a $r^\text{th}$ \emph{residue} modulo $n$ if there exists an integer $\symbLadder \in \ZnZ{n}$ such that $\symbLadder^r = a \bmod n$, otherwise it is a \emph{non-residue} modulo $n$.
\end{defi}

\begin{lem}[Gauss \cite{Gauss65}]
\label{theo:Gauss}
Let $p$ be a prime and $r \ge 2$ be an integer.
We denote $b = \pgcd{p - 1}{r}$.
We have:
    \begin{enumerate}
        \item For every integer $a \in \ZnZinv{p}$, $a$ is a $r^\text{th}$ residue modulo $p$ if and only if $a^{\frac{p - 1}{b}} = 1 \bmod p$,
        \item there exists exactly $\frac{p - 1}{b}$ $r^\text{th}$ residues in $\ZnZinv{p}$,
        \item if $a \in \ZnZinv{p}$ is a $r^\text{th}$ residue modulo $p$ then there exists exactly $b$ integers $\symbLadder \in \ZnZ{p}$ such that $\symbLadder^r = a \bmod p$.
    \end{enumerate}
\end{lem}

According to the chinese remainder theorem, $\symbLadder^3 - a$ is invertible modulo $n = pq$ if and only if $\symbLadder^3 - a$ is invertible modulo $p$ and $\symbLadder^3 - a$ is invertible modulo $q$.
Moreover, because $p$ is prime, $\symbLadder^3 - a$ is invertible $\bmod$ $p$ if and only if $\symbLadder^3 - a \not= 0 \bmod p$, i.e. $a$ is a $3^\text{rd}$ non-residue modulo $p$.
Therefore, the constraint \condN{4} is not satisfied only if $a$ is a $3^\text{rd}$ residue modulo $p$ or $q$.

For $a$ fixed, because $b_p = \pgcd{p - 1}{3} = 1$ or $3$ and $b_q = \pgcd{q - 1}{3} = 1$ or $3$, according to \Cref{theo:Gauss} there are at most $9$ non invertible $\symbLadder^3 - a$ integers modulo $n$, so the probability to pick randomly a ladder constant in $\intervalLadder{a}{n}$ not verifying \condN{4} is:
\begin{align}
    \proba{}{\text{not } \condN{4}}
        &\le \frac{9}{n - 4}\nonumber
\end{align}

In the worst case, the unfavorable cases are distinct, thus the probability to pick randomly a suitable ladder constant is:
\begin{align}
    \proba{}{\condN{1}\condN{2}\condN{3}\condN{4}}
        &\ge 1 - \left(\frac{p + q - 2}{n - 4} + \frac{2}{n - 4} + \frac{9}{n - 4}\right)\nonumber\\
        &= 1 - \frac{p + q + 9}{n - 4} \approx 1 - \frac{1}{p} - \frac{1}{q} - \frac{1}{n}\nonumber
\end{align}
which is almost equal to $1$ in a cryptographic context.
Thus, in the RSA case $n = pq$ finding a suitable ladder constant (Lines~\ref{algo:line:ladderConstant-begin}--\ref{algo:line:ladderConstant-end}) in \Cref{table:strongExample} costs only one iteration with probability almost equal to $1$.

\subsection{Fully-Interleaved Ladder (DSA Case)}
\label{sec:exponentiation:DSA}

In Digital Signature Algorithm (DSA) \cite{NIST186-4} context (or for Diffie-Hellman key exchange mechanism \cite{DH99}), the exponentiation is computed modulo $n = p$, where $p$ is prime.

As in the RSA case we prove that finding a suitable ladder constant costs only one iteration with probability almost equal to $1$.
The argument is similar to the previous one but even simpler.
If $\symbLadder \in \intervalLadder{a}{n} = [2, n - 2] \setminus \set{a}{}$ then \condN{1} $\symbLadder \not= \cThenA$.
Moreover $\symbLadder \not= 0 \bmod n$ so \condN{2} $\symbLadder$ is invertible modulo $n$.
Finally, $\symbLadder \not= \pm 1$ so \condN{3} $\symbLadder^2 - 1 \not= 0 \bmod n$ thus $\symbLadder^2 - 1$ is invertible modulo $n$.

Remains the last constraint \condN{4} $\symbLadder^3 - \cThenA$ is invertible modulo $n$.
In the following we denote $\residue{n}{a}$ that $a$ is a cubic residue modulo $n$, and let $b=\text{pgcd}(n-1,3)$.

According to \Cref{theo:Gauss}, there exists $\frac{n - 1}{b}$ cubic residue in $\ZnZinv{n}$.
$-1$ and $1$ are cubic residue modulo $n$, as opposed to $0$.
Thus, the probability that an integer $a \in [2, n - 2]$ is a cubic residue is:
\begin{align}
    \proba{}{\residue{n}{a}}
        &= \frac{\frac{n - 1}{b} - 2}{n - 3}
        = \frac{1}{b} - \frac{2\left(1 - \frac{1}{b}\right)}{n - 3}\nonumber
\end{align}

According to \Cref{theo:Gauss}, if $a$ is a cubic residue then there exists $b$ cubic roots $\symbLadder$ of $a$ in $\ZnZinv{n}$.
If $\symbLadder = 0, \pm 1$ then $a = \symbLadder^3 = 0, \pm 1$, which is excluded.
If $\symbLadder = a$ then $a = a^3$ i.e. $a(a + 1)(a - 1) = 0 \bmod p$, thus $a = 0, \pm 1$ which is excluded.
So the cubic roots $\symbLadder$ of $a$ are automatically in $\intervalLadder{a}{n} = [2, n - 2] \setminus \set{a}{}$.
Thus the probability that an integer $\symbLadder \in \intervalLadder{a}{n}$ is a cubic root of $a$ is $\frac{b}{n - 4}$.
Therefore, if $a$ is a cubic residue then the probability that $\symbLadder \in \intervalLadder{a}{n}$ satisfies the last constraint is:
\begin{align}
    \proba{}{\condN{4} \knowing \residue{n}{a}}
        &= 1 - \frac{b}{n - 4}\nonumber
\end{align}

Finally, if $a$ is not a cubic residue then every $\ell \in \intervalLadder{a}{n}$ satisfies \condN{4}.
Therefore, the probability to pick randomly a suitable ladder constant is:
\begin{align}
    \proba{}{\condN{1}\condN{2}\condN{3}\condN{4}}
        &= \proba{}{\condN{4}}\nonumber\\
        &= \proba{}{\condN{4} \knowing \residue{n}{a}} \times \proba{}{\residue{n}{a}}\nonumber\\
        &\quad + \proba{}{\condN{4} \knowing \text{not } \residue{n}{a}} \times \proba{}{\text{not } \residue{n}{a}}\nonumber\\
        &= \left(1 - \frac{b}{n - 4}\right)\left(\frac{1}{b} - \frac{2\left(1 - \frac{1}{b}\right)}{n - 3}\right)\nonumber\\
        &\quad + 1 \left(1 - \frac{1}{b} + \frac{2\left(1 - \frac{1}{b}\right)}{n - 3}\right)\nonumber\\
        &= 1 - \frac{1}{n - 4} + \frac{2(b - 1)}{(n - 3)(n - 4)}\nonumber
\end{align}
%
which is almost equal to $1$ in a cryptographic context.

\section{Scalar Multiplication in ECC}
\label{sec:scalarMult}

As in \Cref{sec:exponentiation} we demonstrate concrete examples of the ladder equations, this time for the scalar multiplication used in elliptic curve cryptography (ECC).

ECC was independently introduced in 1985 by Neal Koblitz \cite{Kob87} and Victor Miller \cite{Mil85}. It is nowadays considered as an excellent choice for key exchange or digital signatures, especially when these mechanisms run on resource-constrained devices.
The security of most cryptocurrencies is based on ECC, which has been standardized by the NIST \cite{BCR+18,NIST186-4}.

\begin{defi}[Elliptic Curve]
\label{defi:EC}
Let $p$ be a prime.
An elliptic curve in short Weierstrass form over a finite field $\Field{p}$ is defined by the set $E(\Field{p}) = \set{(x,y) \in \Field{p} \times \Field{p}}{y^2 = x^3+ax+b} \cup \pointO,$ with $a,b \in \Field{p}$ satisfying $4a^3 + 27b^2 \neq 0$ and $\pointO$ being called the point at infinity.
\end{defi} 

The set $E(\Field{p})$ is an additive abelian group with an efficiently computable
group law.
Point addition $P + Q$ or point doubling $2P$ involves additions and multiplications over $\Field{p}$. The point $\pointO$ is the identity element of the group law.
Depending on the parameters $a$ and $b$ of the curve, there exists many formulas to optimize these two operations.
A large and updated survey can be found on the elliptic curves explicit formulas database web site:
\url{https://www.hyperelliptic.org/EFD/}

The main operation in ECC is scalar multiplication $k\pointA = \pointA + \dots + \pointA$, where $\pointA$ is a
point on a curve and $k$ is an integer.
It can be performed by using the \emph{double-and-add} algorithm, similar to the square-and-multiply algorithm in \Crpref{table:squareMultiplyAlways}. 
The initialization $x \leftarrow 1$ is replaced by $\pointP \leftarrow \pointO$,
the squaring $x \leftarrow x^2$ is replaced by a doubling $\pointP \leftarrow 2\pointP$
and the multiplication $x \leftarrow a x$ is replaced by an addition $\pointP \leftarrow \pointA + \pointP$,
producing the output $\pointP = k\pointA$.

Thus, for a fixed point $\pointA$ we consider in this \namecref{sec:scalarMult} the following known functions:
\begin{align}
    \funThen{\pointP} &= 2\pointP + \pointA\nonumber\\
    \funElse{\pointP} &= 2\pointP\nonumber
\end{align}
And we consider ladder functions of the following form:
\begin{align}
    \funFirst{\pointP}{\pointQ} &= \cFirstPointP\pointP + \cFirstPointQ\pointQ + \cFirstPointA\pointA\nonumber\\
    \funSecond{\pointP}{\pointQ} &= \cSecondPointP\pointP + \cSecondPointQ\pointQ + \cSecondPointA\pointA\nonumber\\
    \funLadder{\pointP} &= \cLadderPointP\pointP + \cLadderPointA\pointA\nonumber
\end{align}
with $\cFirstPointP, \cFirstPointQ, \cLadderPointP \not= 0$ for both cases, and $\cSecondPointP, \cSecondPointQ \not= 0$ for the fully-interleaved ladders case,
allowing in principle non-secure scalar multiplications as intermediate computations for the secure one.
The efficiency of the obtained solutions are discussed in \Cref{sec:scalarMult:ECC}.

\subsection{Both Semi- and Fully-Interleaved Ladders}
\label{sec:scalarMult:commonEqs}

\Cref{eq:weak:thenValue} (resp. \Cref{eq:strong:thenValue}) $\funFirst{x}{\funLadder{x}} = \funThen{x}$ and \Cref{eq:weak:elseLadder} (resp. \Cref{eq:strong:elseLadder}) $\funFirst{\funLadder{x}}{x} = \funLadder{\funElse{x}}$ for the semi- (resp. fully-) interleaved cases imply that:
\begin{align}[left=\empheqlbrace]
    \cFirstPointP + \cFirstPointQ\cLadderPointP
        &= 2\nonumber\\
    \cFirstPointQ
        &= (2 - \cFirstPointP)\cLadderPointP\nonumber
\end{align}
So $(2 - \cFirstPointP)(\cLadderPointP^2 - 1) = 0$.
If $\cFirstPointP = 2$ then $\cFirstPointQ = 0$, thus we assume $\cFirstPointP \not= 2$.
To make no assumption about the finite field, we assume for the following $\cLadderPointP = \pm 1$.
They also imply that:
\begin{align}[left=\empheqlbrace]
    \cFirstPointQ\cLadderPointA + \cFirstPointA
        &= 1\nonumber\\
    \cFirstPointA
        &= (2 - \cFirstPointP)\cLadderPointA - \cLadderPointA\nonumber
\end{align}
So $(2 - \cFirstPointP)(\cLadderPointP + 1)\cLadderPointA = 1 + \cLadderPointA$.

\begin{itemize}
\item If $\cLadderPointP = -1$ then $\cLadderPointA = -1$, $\cFirstPointQ = \cFirstPointP - 2$ and $\cFirstPointA = (\cFirstPointP - 2) + 1$, thus the candidate ladder functions are:
\begin{align}
    \funFirst{\pointP}{\pointQ} &= \cFirstPointP(\pointP + \pointQ + \pointA) - (2\pointQ + \pointA)\nonumber\\
    \funLadder{\pointP} &= - (\pointP + \pointA)\nonumber
\end{align}

\item If $\cLadderPointP = 1$ then by assuming that $3 - 2\cFirstPointP$ is invertible we obtain $\cLadderPointA = \frac{1}{3 - 2\cFirstPointP}$, $\cFirstPointQ = 2 - \cFirstPointP$ and $\cFirstPointA = 1 - \frac{2 - \cFirstPointP}{3 - 2\cFirstPointP}$, thus the candidate ladder functions are:
\begin{align}
    \funFirst{\pointP}{\pointQ} &= \cFirstPointP(\pointP - \pointQ) + 2\pointQ + \frac{1 - \cFirstPointP}{3 - 2\cFirstPointP}\pointA\nonumber\\
    \funLadder{\pointP} &= \pointP + \frac{1}{3 - 2\cFirstPointP}\pointA\nonumber
\end{align}
\end{itemize}

\subsection{Semi-Interleaved Ladder}
\label{sec:scalarMult:weakCase}

The remaining \Cref{eq:weak:thenLadder} $\funElse{\funLadder{x}} = \funLadder{\funThen{x}}$ for the semi-interleaved ladders is already satisfied for the $\cLadderPointP = -1$ case.
For the $\cLadderPointP = 1$ case, it implies that $(\frac{1}{3 - 2\cFirstPointP} - 1)\pointA = 0$, thus to avoid constraint on $\pointA$ we assume $\cFirstPointP = 1$.

\begin{theo}
\label{theo:multSIL}
The double-and-add algorithm with $\funThen{\pointP} = 2\pointP + \pointA$ and $\funElse{\pointP} = 2\pointP$ is semi-ladderizable with two solutions:
\begin{enumerate}
\item
\begin{align}
    \funFirst{\pointP}{\pointQ} &= \cFirstPointP(\pointP + \pointQ + \pointA) - (2\pointQ + \pointA)\nonumber\\
    \funLadder{\pointP} &= - (\pointP + \pointA)\nonumber
\end{align}
where $\cFirstPointP \not= 0, 2$, and:
\item
\begin{align}
    \funFirst{\pointP}{\pointQ} &= \pointP + \pointQ\nonumber\\
    \funLadder{\pointP} &= \pointP + \pointA\nonumber
\end{align}
\end{enumerate}
\end{theo}
The second solution is actually the common Montgomery ladder for the scalar multiplication.
The first one for $\cFirstPointP = 1$ corresponds to the ladder function $\funFirst{\pointP}{\pointQ} = P - Q$ which is similar to the common Montgomery ladder except that the point $\pointQ$ has an opposite sign.

\subsection{Fully-Interleaved Ladder}
\label{sec:scalarMult:strongCase}

In the case $\cLadderPointP = -1$, \Cref{eq:strong:thenLadder} $\funSecond{\funThen{x}}{\funLadder{x}} = \funLadder{\funThen{x}}$ for the fully-interleaved ladders is equivalent to:
\begin{align}[left=\empheqlbrace]
    2\cSecondPointP - \cSecondPointQ
        &= -2\nonumber\\
    \cSecondPointP - \cSecondPointQ + \cSecondPointA
        &=-2\nonumber
\end{align}
So $\cSecondPointQ = 2(1 + \cSecondPointP)$ and $\cSecondPointA = \cSecondPointP$.
Thus \Cref{eq:strong:elseValue} $\funSecond{\funFirst{\funLadder{x}}{x}}{x} = \funElse{x}$ for the fully-interleaved ladders implies that $\cFirstPointP\cSecondPointP = 0$.
Because we assumed $\cFirstPointP, \cSecondPointP \not= 0$ and we are looking for a solution independent from the chosen field for the coefficients, we have no solution for this case.

In the case $\cLadderPointP = 1$, \Cref{eq:strong:thenLadder} $\funSecond{\funThen{x}}{\funLadder{x}} = \funLadder{\funThen{x}}$ for the fully-interleaved ladders is equivalent to:
\begin{align}[left=\empheqlbrace]
    2\cSecondPointP + \cSecondPointQ
        &= 2\nonumber\\
    \cSecondPointP + \frac{\cSecondPointQ}{3 - 2\cFirstPointP} + \cSecondPointA
        &= 1 + \frac{1}{3 - 2\cFirstPointP} \nonumber
\end{align}
So $\cSecondPointQ = 2(1 - \cSecondPointP)$ and $\cSecondPointA = 1 + \frac{(2\cFirstPointP - 1)\cSecondPointP - 1}{3 - 2\cFirstPointP}$.
Thus \Cref{eq:strong:elseValue} $\funSecond{\funFirst{\funLadder{x}}{x}}{x} = \funElse{x}$ for the fully-interleaved ladders implies that $\frac{2(\cFirstPointP(\cSecondPointP - 1) + 1)}{3 - 2\cFirstPointP}\pointA = 0$.
To avoid constraint on $\pointA$ we assume $\cFirstPointP(\cSecondPointP - 1) + 1 = 0$, i.e. $\cSecondPointP = 1 - \frac{1}{\cFirstPointP}$.

\begin{theo}
\label{theo:multFIL}
The double-and-add algorithm with $\funThen{\pointP} = 2\pointP + \pointA$ and $\funElse{\pointP} = 2\pointP$ is fully-ladderizable with:
\begin{align}
    \funFirst{\pointP}{\pointQ} &= \cFirstPointP\left(\pointP - \pointQ - \frac{1}{3 - 2\cFirstPointP}\pointA\right) + 2\pointQ + \frac{1}{3 - 2\cFirstPointP}\pointA\nonumber\\
    \funSecond{\pointP}{\pointQ} &= \pointP + \frac{1}{\cFirstPointP}\left(- \pointP + 2\pointQ + \frac{1}{3 - 2\cFirstPointP}\pointA\right) - \frac{1}{3 - 2\cFirstPointP}\pointA\nonumber\\
    \funLadder{\pointP} &= \pointP + \frac{1}{3 - 2\cFirstPointP}\pointA\nonumber
\end{align}
where $\cFirstPointP \not= 0, 1, 2$, and where $\cFirstPointP$ and $3 - 2\cFirstPointP$ are invertible.
\end{theo}
The ladder functions of this solution for $\cFirstPointP = 1$ are $\funFirst{\pointP}{\pointQ} = \pointP + \pointQ$, $\funSecond{\pointP}{\pointQ} = 2\pointQ$ and $\funLadder{\pointP} = \pointP + \pointA$, i.e. the common Montgomery ladder for the scalar multiplication, but is no longer a fully-interleaved ladder.

\subsection{Application to ECC}
\label{sec:scalarMult:ECC}

The novel candidates are the first semi-interleaved ladder in \Cref{theo:multSIL} requiring a scalar multiplication by $\cFirstPointP$, and the fully-interleaved ladder in \Cref{theo:multFIL} requiring a scalar multiplication by $\frac{1}{3 - 2\cFirstPointP}$ then $\cFirstPointP$ and $\frac{1}{\cFirstPointP}$.

Scalar multiplications $c\pointA$ for various coefficients $c$ can be precomputed if $\pointA$ was a fixed point, as in the DSA signing step or the first step of Diffie-Hellman protocol.
But both candidates involve scalar multiplications for non-fixed points $\pointR$, linear combination of $\pointA$, $\pointP$, and $\pointQ$.


A scalar multiplication $c \pointR$ for a specific coefficient $c$ can be efficiently computed (i.e. with very few field operations) by using an elliptic curve endowed with an efficient endomorphism.
On such curves, there exists a constant $\cEndormophism$, depending on the modulus $p$ and the parameters of the curve (see \Cref{defi:EC}), such that $\cEndormophism R$ can be computed with very few multiplications (only one in most cases).
This is detailed in \cite{GalLamVan01-FasterPoint}, where the authors describe the GLV method used in various standards, like the Bitcoin protocol specification or TLS \cite{TLS20}.

Another way to optimize the cost of $c R$ is to use a short addition chain \cite{KNU97}, so that the number of point operations to compute $c R$ is equal to the length of the chain.
The problem of computing the shortest addition chain for a given integer is hard \cite{BAH06} but for ``small'' integers there exists tables giving the corresponding chain, like \url{https://bo.blackowl.org/random/ln}.

Therefore, for the first semi-interleaved ladder in \Cref{theo:multSIL}, one can use an elliptic curve endowed with an efficient endomorphism $\cEndormophism$ and assign $\cFirstPointP = \cEndormophism$.
Or, for a non-specific curve, one can choose $\cFirstPointP$ amongst the integers corresponding to ``very short'' addition chains.
In that case, because as in \Cref{theo:expSIL} $\cFirstPointP$ does not occur in $\funLadder{\pointP}$, $\cFirstPointP$ can be chosen randomly for every iteration, thus increasing security compared to the common Montgomery ladder.
However, this comes at the price of drastically increasing the computational cost of the scalar multiplication.

For the fully-interleaved ladder candidate, neither an efficient endomorphism nor an addition chain can be used.
Indeed, that $\cFirstPointP\pointR$ can be efficiently computed does not imply that $\frac{1}{\cFirstPointP}\pointR$ can also be efficiently computed, because both coefficients are dependent.
Therefore, the fully-interleaved ladder candidate might not be applicable in practice to ECC.

\section{Conclusion and Future Work}
\label{sec:conclu}

\label{sec:conclu:summary}

We abstracted away the algorithmic strength of the Montgomery ladder against side-channel and fault-injection attacks, by defining semi- and fully-ladderizable algorithms.
We designed also fault-injection attacks able to obtain some/all bits of the secret key from the semi-interleaved ladders like the Montgomery ladder, and even from the fully-interleaved ladders if the attacker is able to stuck-at the key register.

As examples, we provided for the modular exponentiation a better semi-interleaved ladder using a random mask updated at every iteration, and a fully-interleaved ladder depending on an appropriate ladder constant.
We investigated the properties required by the ladder constant, and discovered that they are almost always satisfied in a cryptographic context.

To demonstrate that generality of our approach, we provided new algorithms for the scalar multiplication in elliptic curves: a novel and effective semi-interleaved ladder, and a fully-interleaved ladder.
\label{sec:conclu:futureWork}
The applicability of the last candidate depends on the efficient computations of several dependent coefficients for the scalar multiplications, which is an open and interesting problem.

Finally, we only investigated the univariate case for conditional branching, but the multivariate case may be of interest.
For instance, a multi-variable polynomial $\sum_{0 \le n \le d}\sum_{n_1 + \dots + n_k = n}c_{n_1,\dots,n_k}\prod_{1 \le i \le k}x_i^{n_i}$ can be represented by a multidimensional array of coefficients, thus the manipulation of ladder equations could be handled by using matrix operations.


\section*{Acknowledgments}
\label{sec:conclu:acknowledgement} 

The authors want to thank the reviewers of ARITH 2020 for their relevant and very useful comments.

\bibliographystyle{IEEEtran}
\bibliography{IEEEabrv,bibliography}

\end{document}